\documentclass[aps,prl,twocolumn,superscriptaddress]{revtex4-2}

\usepackage{graphicx}
\usepackage{amsmath,amssymb}
\usepackage{color}
\usepackage{bm,bbm,bbold}
\usepackage{multirow}

\usepackage[bookmarks=true,colorlinks=true,urlcolor=blue,linkcolor=blue,citecolor=blue,breaklinks]{hyperref}

\usepackage{booktabs}

\usepackage{color}
\hypersetup{
	colorlinks   = true,
	citecolor    = blue
}

\usepackage{ulem}

\usepackage{multirow}
\usepackage{siunitx} % use this package module for SI units

\usepackage{xspace}

\newcommand{\bfk}{\mathbf{k}}

\newcommand{\bfr}{\mathbf{r}}
\newcommand{\vect}[1]{\mathbf{#1}}

\newcommand{\abs}[1]{\left\lvert #1 \right\rvert}
\newcommand{\figref}[1]{Fig.~\ref{#1}}
\newcommand{\eqnref}[1]{Eq.~(\ref{#1})}

\newcommand{\moire}[0]{moir\'{e}\xspace}

\newcommand{\vF}[0]{v_\mathrm{F}}
\newcommand{\tbgmagic}[0]{\theta^{\mathrm{M}}_\mathrm{tBG}}

\definecolor{pu}{RGB}{200,50,200}
\definecolor{gr}{RGB}{0,187,0}
\definecolor{bl}{RGB}{68,34,200}
\definecolor{re}{RGB}{200,34,68}
\definecolor{ye}{RGB}{255,165,0}
\definecolor{oran}{RGB}{255,170,0}

\begin{document}
	
	%\title{Ubiquitous magic angles in generalized trilayer graphene}
	\title{Extended magic phase in twisted graphene multilayers}
	
	\author{D. C. W. Foo}
	\email{c2ddfcw@nus.edu.sg}
	\affiliation{Centre for Advanced 2D Materials, 
		National University of Singapore, 6 Science Drive 2, Singapore 117546}
	\author{Z. Zhan}
	\affiliation{IMDEA Nanociencia, C/ Faraday 9, 28049 Madrid, Spain}
	\author{Mohammed M. Al Ezzi}
	\affiliation{Centre for Advanced 2D Materials, 
		National University of Singapore, 6 Science Drive 2, Singapore 117546}
	\affiliation{Department of Physics, Faculty of Science, 
		National University of Singapore, 2 Science Drive 3, Singapore 117542}
	\author{L. Peng}
	\affiliation{Department of Physics, Faculty of Science, 
		National University of Singapore, 2 Science Drive 3, Singapore 117542}
	%\author{G. Vignale}
	%\affiliation{The Institute for Functional Intelligent Materials, National University of Singapore, 4 Science Drive 2, Singapore 117544}
	\author{S. Adam}
	\affiliation{Centre for Advanced 2D Materials, 
		National University of Singapore, 6 Science Drive 2, Singapore 117546}
	\affiliation{Department of Physics, Faculty of Science, 
		National University of Singapore, 2 Science Drive 3, Singapore 117542}
	\affiliation{Department of Materials Science and Engineering, 
		National University of Singapore, 9 Engineering Drive 1, 
		Singapore 117575}
	\affiliation{Yale-NUS College, 16 College Ave West, Singapore 138527}
	\author{F. Guinea}
	\affiliation{IMDEA Nanociencia, C/ Faraday 9, 28049 Madrid, Spain}
	\affiliation{Donostia International Physics Center, Paseo Manuel de Lardiz\'abal 4, 20018, San Sebasti\'an, Spain}

	\begin{abstract}
		Theoretical and experimental studies have verified the existence of ``magic angles'' in twisted bilayer graphene, where the twist between layers gives rise to flat bands and consequently highly correlated phases.  Narrow bands can also exist in multilayers with alternating twist angles, and recent theoretical work suggests that they can also be found in trilayers with twist angles between neighboring layers in the same direction. 
		We show here that flat bands exist in a variety of multilayers where the ratio between twist angles is close to coprime integers. We generalize previous analyses, and, using the chiral limit for interlayer coupling, give examples of many combinations of twist angles in stacks made up of three and four layers which lead to flat bands. The technique we use can be extended to systems with many layers. Our results suggest that flat bands can exist in graphene multilayers with angle disorder, that is, narrow samples of turbostatic graphite.

		%An extension to graphene multilayers was subsequently proposed and experimentally verified up to at least $n=6$ layers where, provided the twist angles were fixed in magnitude but alternating in sign, a new series of magic angles emerge with a simple geometric relationship between the bilayer magic angle and the multilayer magic angle.  Further work found numerical flat bands in equal magnitude, equal sign trilayers, and strongly correlated phases in generically twist trilayers.  We explore the space of generic trilayers and find a generalised, extended magic phase, recovering exactly the known, isolated points and accurately reproducing experimental observables.  We further more demonstrate the applicability of our methods to stacks of more graphene layers.  We argue that many of the magic configurations we propose should be more robust to strain relaxation \textcolor{red}{and Hartree band corrections than other known flat bands.}
	\end{abstract}
	
	\maketitle
	
	{\it Introduction }\label{sec:intro} -- Twisted bilayer graphene (tBG) has proven to be a fertile platform for condensed matter physics studies owing to the remarkable tunability of the Fermi velocity and electronic structure through the twist angle~\cite{LPN07}.  In particular, at the so-called ``magic angles''~\cite{BM11} for tBG, $\tbgmagic\approx1.1^\circ$, the Fermi velocity vanishes, allowing for strongly correlated phases where interactions dominate over the kinetic energy~\cite{AM20,BDEY20}. 
	
	This has led to a resurgence of interest in \moire systems, including but not limited to Van der Waals heterostructures~\cite{CL21,Zetal21,KL22,SS23}, transition metal dichalcogenides~\cite{Letal21,tmd2,VCF22,EFMF22}, and of course graphene multilayers~\cite{CKLR22,TPB22,Letal22,Betal22}.  One particularly successful avenue of investigation has been the extension of the arguments demonstrating the existence of magic angles in tBG~\cite{SGG12,TKV19} to graphene multilayers~\cite{KKTV19}, which found simple geometric relations between the bilayer and multilayer magic angles, predictions that were later verified experimentally~\cite{Cetal21,Ketal22,Petal22}.  Later numerical studies have built upon these foundations, demonstrating for example that the existence of flat bands is robust against perpendicular magnetic fields~\cite{SS21} and asymmetry of the layer parameters~\cite{EYFS22}, and that peaks in the density of states (DOS) appear, and are resistant to relaxation, for trilayer configurations beyond the canonical alternating-twist case~\cite{MZY23}.  Recent work has predicted magic angles in the equal twist case~\cite{PT23}, and a magic line, with magic angle a function of pressure, has been predicted in twisted WSe\textsubscript{2}~\cite{Metal23}. In this work we demonstrate the existence of flat bands in multilayers with generic, commensurate twist angles, indicating an extended $N-2$ dimensional magic surfaces in the space of $N-1$ twist angles characterising $N$ layers, with particular emphasis on generic twisted trilayer graphene (tTG).
	
	A prior study~\cite{KKTV19} noted that in the particular case of unshifted, alternating twist graphene multilayers, the Hamiltonian may be exactly decomposed into a direct sum of bilayer graphene Hamiltonians (with an additional monolayer for odd number of layers), with the \moire potential identical up to a scalar multiple.  This allows one to directly read off the magic angles of multilayer stacks.  In contrast, other trilayer configurations have more complex effective \moire potentials, though magic angle physics may still be possible~\cite{LEFS22}.  In particular, a strongly correlated superconducting state has been observed in a trilayer system with incommensurate twist angles~\cite{Uetal23}. 
	
\begin{figure}[t]
		\centering
		%\begin{tabular}{lr}
		\includegraphics[width=0.5\textwidth]{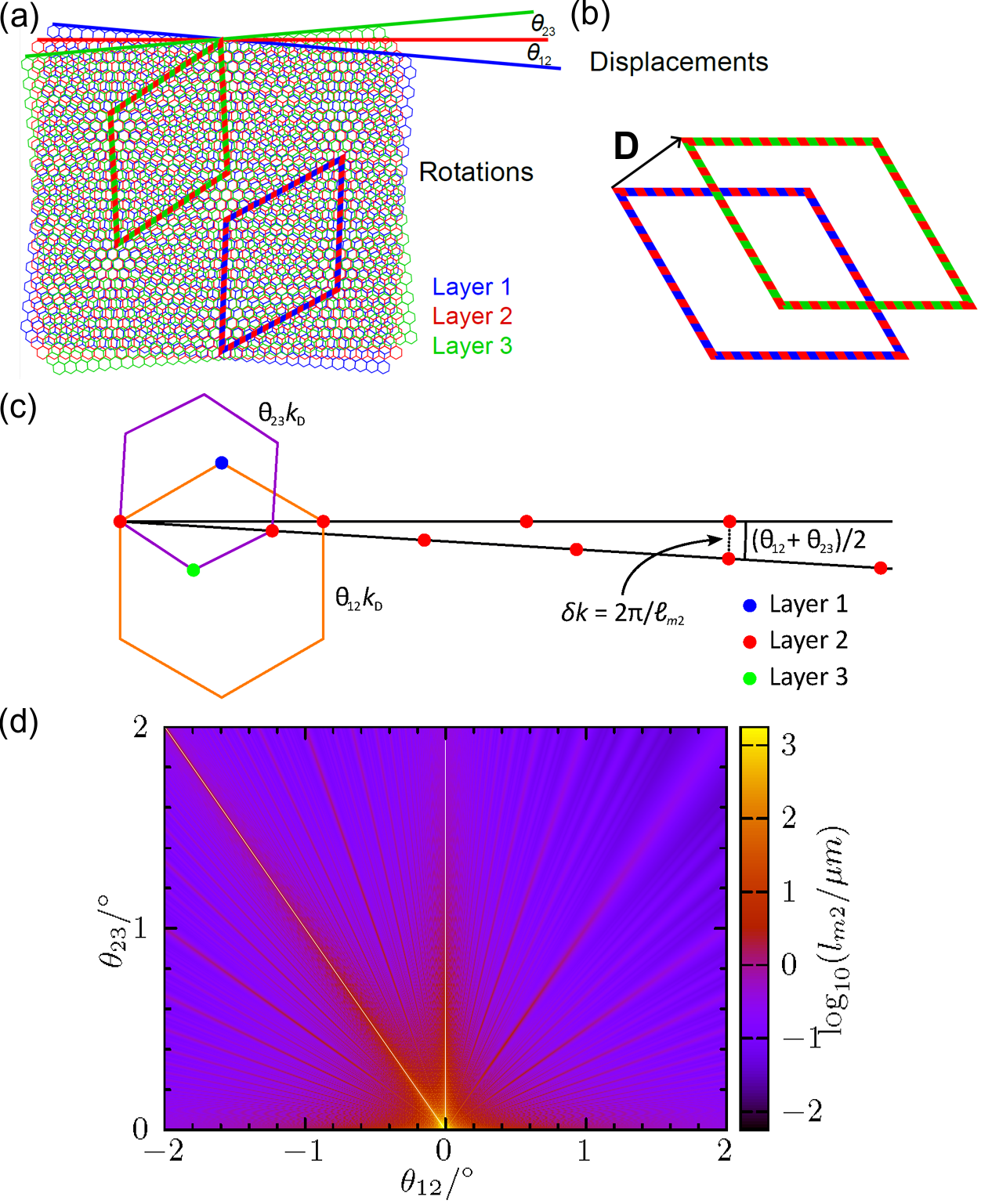} 
		%\includegraphics[width=0.25\linewidth]{dos_points_trilayer_11.jpg}
		%\end{tabular}
		%\includegraphics[width=0.60\linewidth]{dos_trilayer_34.jpg}
		\caption{Schematic illustrating the trilayer parameters (a) twist angle pairs and (b) the \moire displacement $\mathbf{D}$. The moir\'e unit cell of tBG with $\theta_{12}$ is outlined by a rhombus with alternating colors red and blue, and $\theta_{23}$ with colors red and green. The rhombus vertices lie at $AA$ stacking regions of the respective layers. (c) Sketch of connected momentum states between layers with misaligned \moire patterns (orange and purple), resulting in a \moire-of-\moire modulation. Large angles $\theta_{12}=4^\circ$, $\theta_{23}=3^\circ$ are used for clarity, and less misalignment is expected for real magic angle samples. The lengthscale of this modulation is determined by the closest approach of second layer lattice sites formed from the two \moire patterns, indicated with a dotted line. The local expansion used in numerical calculation neglects this misalignment and, for commensurate angles, collapses this distance to zero. (d) Log-scale plot of the \moire-of-\moire length scale, $l_{m2}$, in microns, as a function of twist angles. The left diagonal in white marks the line $\theta_{12}=-\theta_{23}$, where there is no misalignment and $l_{m2}$ diverges. $l_{m2}$ also diverges when either angle is zero and there is no \moire-of-\moire pattern to speak of.}
		\label{fig:moire}
	\end{figure}

	%%%%%%%%%%%%%%%%%%%%%%%%%%%%%%%%%%%%%%%%%%%
	{\it Moir\'e lattices in twisted graphene trilayers and in other twisted stacks.} --
	The electronic structure of a twisted bilayer can be reasonably approximated by a periodic moir\'e structure~\cite{LPN07,BM11}, irrespective of whether the two layers are commensurate at the atomic scale. The situation is more complex in twisted stacks with more than two layers~\cite{K15,AC18,CWG19,LWM19,MRB19,WZS20,ZCMLK20,MZY23}. In a trilayer, for instance, two twist angles between nearest neighbor layers can be defined, $\theta_{12}$ and $\theta_{23}$, shown in Fig. \ref{fig:moire}(a). Except for the case mentioned above of multilayers where the twist angle is of equal magnitude and opposite sign between nearest neighbor pairs of layers~\cite{KKTV19}, no simple global moir\'e structure can be defined. 
	
	An approximate moir\'e lattice can be defined if the relative angle between the Brillouin Zones defined by the twist angles between different pairs of layers is neglected~\cite{CWG19,MRB19,WZS20}. For instance, in a trilayer with two different twist angles, $\theta_{12} , \theta_{23}$, this approximation becomes exact when $\theta_{12} = \theta_{23}$ in the limit $\theta_{12} \rightarrow 0$, although corrections~\cite{ZCMLK20} are required for physically relevant angles, $\theta_{12} , \theta_{23} \approx 1^\circ$. It is worth noting that this approximation, for a trilayer, has a dependence not only on the angles $\theta_{12} , \theta_{23}$, but also on the relative displacement of the \moire potentials, $\mathbf{D}$ illustrated in Fig. \ref{fig:moire}(b)~\cite{LWM19}. The dependence is periodic on $\mathbf{D}$, with wavelength of the order of the \moire length scales.
	
	It can be shown~\cite{MGM23} that the approximation mentioned above defined by a moir\'e lattice commensurate with the two twist angles and a displacement between the top and bottom layers describes {\it locally} the full problem of the trilayer. For the case of approximately commensurate twist angles, $\theta_{12} \approx m \theta_0 , \theta_{23} \approx n \theta_0$, where $m , n$ are coprime integers, the interlayer tunneling terms in the trilayer continuum Hamiltonian  can be written as containing two periodicities: i) the periodicity, $\ell_m$, of the moir\'e problem obtained by neglecting the angle between the two Brillouin zones defined by $\theta_{12}, \theta_{23}$, of order $d / \theta_0$, where $d$ is the lattice constant of graphene, and ii) a periodicity, $\ell_{m2}$ obtained by modulating the displacement $\mathbf{D}$, over a larger unit cell whose dimensions scale as $d / \theta_0^2$. Each tBG pair in tTG defines a momentum-space \moire unit cell, purple and orange in Fig. \ref{fig:moire}(c), whose misalignment leads to misalignment of the \moire lattices, shown with the black lines and the finite $\delta k$. Neglecting this \moire misalignment forces the black lines to coincide, and for commensurate angles forces $\delta k=0$. $\delta k$ therefore is conjugate to $\ell_{m2}$. For small values of $\theta_0$, shown in Fig. \ref{fig:moire}(d), this scaling behavior implies that $\ell_{m}  \ll \ell_{m2}$, as $\ell_{m2}$ can reach values of microns for $\theta_0 \lesssim 1^\circ$. Further details may be found in~\cite{SI}.
	
	It is worth noting that small and layer-dependent biaxial strains can exactly align the Brillouin zones associated to different twist angles, $\theta_{12}$, $\theta_{23}$, so that $\ell_{m2} \rightarrow \infty$ and the approximation which makes use of a single moir\'e lattice defined by $\ell_m$ becomes exact, see~\cite{Detal23}. These strains imply a small relaxation of the lattice~\cite{Detal23}. This analysis has been extended to some commensurate cases of  trilayers with unequal twist angles~\cite{NKK23}.
	
	%%%%%%%%%%%%%%%%%%%%%%%%%%%%%%
	{\it Results. Flat bands in the chiral limit.} --
	We first analyze numerically the existence of flat bands in chiral model trilayers~\cite{SGG12,TKV19} for different combinations of twist angles, $\theta_{12} , \theta_{23}$ and relative displacements between the top and bottom layers. The chiral limit for graphene trilayers with the same twist angle between nearest layers, $\theta_{12} = \theta_{23}$ has been recently explored in~\cite{PT23,GMM23,Detal23}. The parameters that we use~\cite{Ketal18} in the following are $\gamma_0 = 2.6$ eV (intralayer nearest neighbor hopping), and $t_{AB} = 0.0975$ eV (interlayer hopping). The choice of $\gamma_0$ leads to a Fermi velocity $v_F \approx 840$ km/s.
	
\begin{figure}[t]
		\centering
		\includegraphics[width=0.5\textwidth]{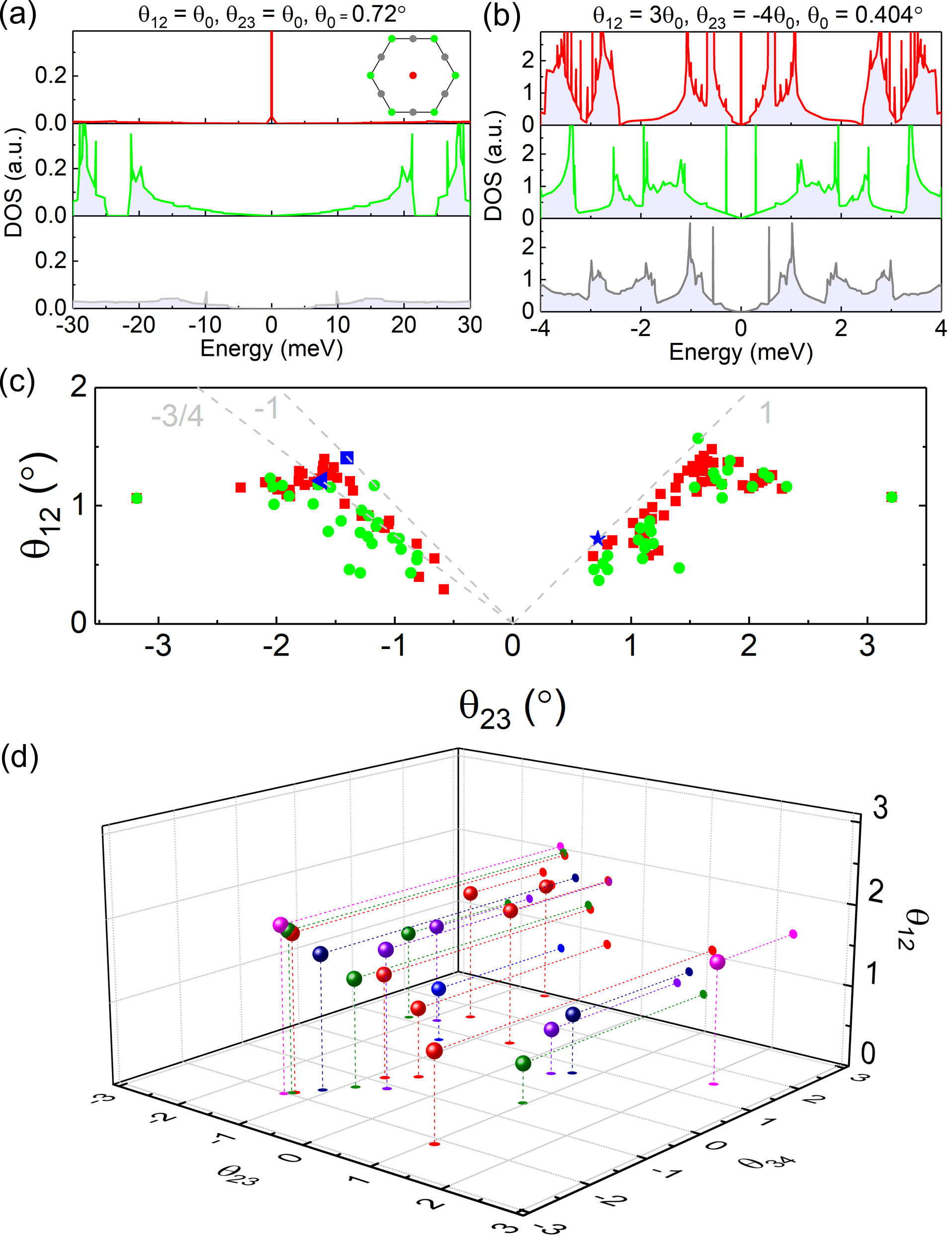}
		\caption{(a) Local DOS of a graphene trilayer in the chiral limit with $\theta_{12} = \theta_{23} =\theta_0= 0.72^\circ$ (magic angle for zero displacement) for three different displacements $\mathbf{D}$ between the top and bottom layers. The insert on the top panel are the three displacements, which are marked by a dot in the hexagon with the color corresponding to that of the local DOS. (b) As in (a), but for a combination of angles $\theta_{12}= 3 \theta_0$, $\theta_{23} = - 4 \theta_0$, and $\theta_0 = 0.404^\circ$. Note the difference in the scales between the plots in (a) and (b). (c) Calculated magic angle pairs ($\theta_{12}$, $\theta_{23}$) for twisted trilayer graphene. The red and green colors are for magic angles with red and green displacements illustrated in insert of (a), respectively. The magic angle pair of (a) is marked by a blue star and that of (b) by a blue triangle. The blue square is the magic angle in twisted trilayer graphene with mirror symmetry. (d) Magic angle triplets ($\theta_{12}$, $\theta_{23}$, $\theta_{34}$) for stacks of four layers. Different colors are used for clarity.}
		\label{magic}
	\end{figure}

	The analytical arguments for the existence of infinitely flat central bands in twisted bilayer graphene~\cite{TKV19} is based on the existence of zeros at the corners of the unit cell of a 2 component spinor which describes the periodic part of the wavefunction at the $K$ point of the Brillouin Zone. As one of the two components, by symmetry, vanishes at the relevant points, the problem reduces to the tuning of the zeros of a single component. The argument can be generalized to a trilayer, where a three component spinor is involved. When the Hamiltonian has ${\cal C}_3$ symmetry~\cite{TKV19} two of the three components of the spinor vanish at the relevant point in the real space unit cell, so that the existence of infinitely flat bands reduces to the existence of a zero in a single component of the spinor. The ${\cal C}_3$ symmetry exists at least for a displacement $\mathbf{D}=0$, and for the displacements which take the trilayer from the $AAA$ to the $ABA$ configuration. An alternative analysis, for the case of a trilayer with $\theta_{12} = \theta_{23}$, can be found in~\cite{PT23}. We numerically find flat bands, whose width is continuously lowered as the angle is determined with better precision, for all combinations of integers, $m, n$ and displacements, $\mathbf{D}$, tried. Results are shown in Fig. \ref{magic}. Interestingly, in some cases we find a quadruplet of bands nearly degenerate at zero energy. 
	We show in Fig. \ref{magic} the local DOS at the magic angles corresponding to $\{m , n \} = \{ 1 , 1 \}$  and $\{m , n \} = \{ 3 , -4 \}$ obtained for a displacement between top and bottom layers $\mathbf{D} = 0$. For each magic angle, we plot the DOS for the selected displacement, $\mathbf{D} = 0$, and for two other displacements, shown in the inset in Fig. \ref{magic} (a). There is a significant dependence of the local DOS on $\mathbf{D}$ in the $\{ 1 , 1 \}$ case, and a less pronounced dependence in the $\{ 3 , - 4 \}$ case. This trend towards a weaker dependence on $\mathbf{D}$ with increasing values of the integers $\{ m , n \}$ is observed for all combinations explored, see~\cite{SI}. 
	
	Alternatively, the $3 \times 3$ matrix equation which describes the spinor at the magic angles of the chiral bilayer~\cite{TKV19,PT23} can be approximately reduced to an effective bilayer, making significantly easier the calculation of magic angles as function of commensuration and displacement, see~\cite{SI}.
	
	\begin{figure}[t!]
		\centering
		%\begin{tabular}{c}
		\includegraphics[width=\linewidth]{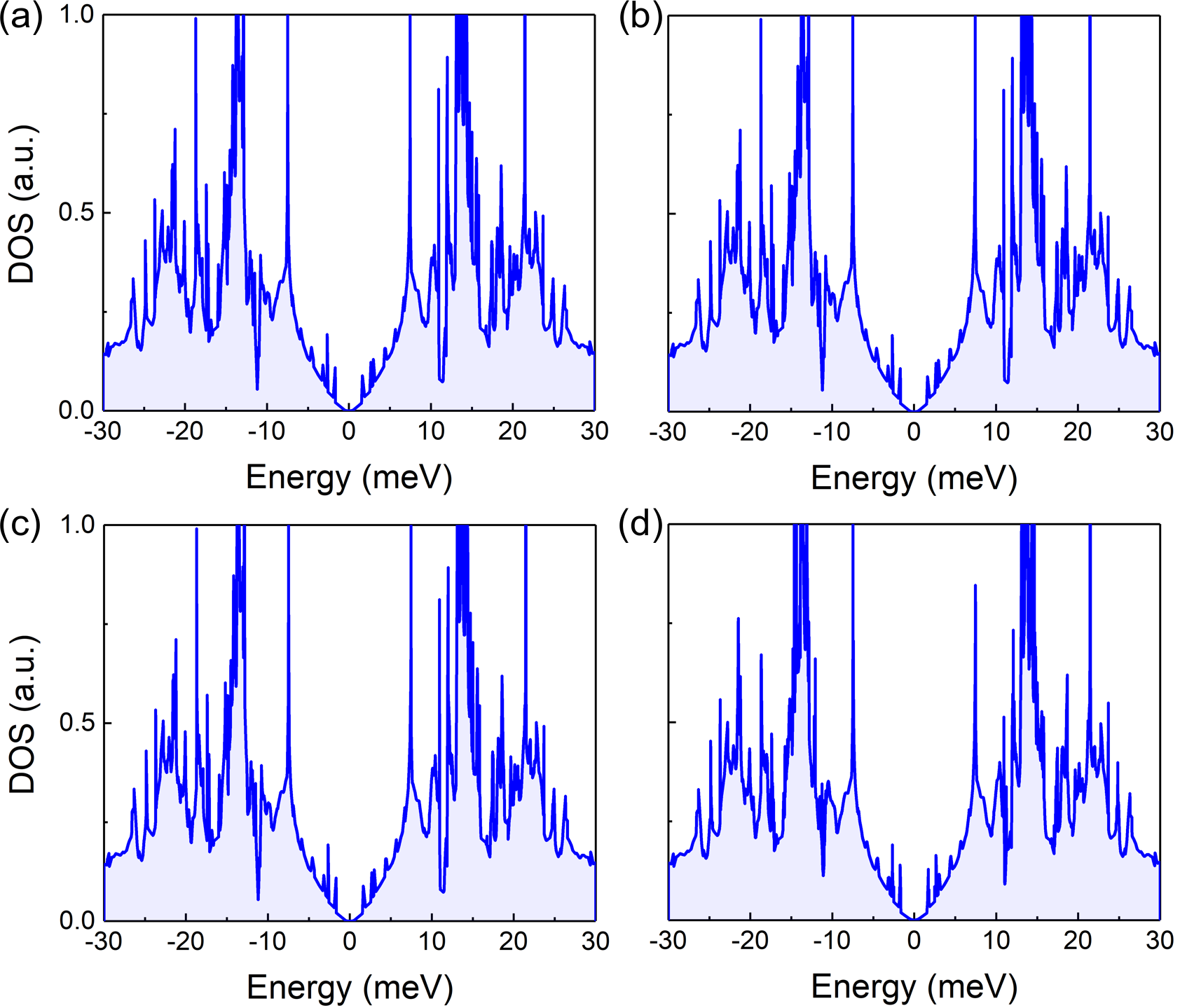} 
		%\includegraphics[width=0.50\linewidth]{dos34.jpg}
		%\end{tabular}
		\caption{Local DOS for a trilayer in the chiral limit with angles $\theta_{12} = 3\theta_0$, $\theta_{23} = -4\theta_0$, $\theta_0= 0.47^\circ$ and different displacements between the top and bottom layers. The displacements are described by the phases $\{ 0 , 0 , 0 \}$ (a), $\{ 0 , - \pi / 18 , \pi / 18 \}$ (b), and $\{ \pi /36 , - \pi / 18 , \pi / 36 \}$ (c). (d) An average of the three densities of states with the weights $\{ 1 : 2 : 3 \}$, is shown in (d).}
		\label{fig:dos34}
	\end{figure}
	
	The scheme discussed here for a trilayer~\cite{CWG19,MRB19,WZS20,MGM23}, based on two twist angles, two integers, and one interlayer displacement can be extended to an $N$ layer model, with $N-1$ twist angles, and $N-2$ independent displacements. For the case of $N-$layer stacks where the twist angles between neighboring layers are equal, an analytical expression for the first magic angle, which generalizes the results in~\cite{PT23}, gives:
	\begin{align}
		\theta_N^M&= \theta_2^M \left[ \frac{1}{(N-1)!} \right]^{\frac{1}{N-1}}
	\end{align}
	This expression gives, for $N \rightarrow \infty$
	%\begin{align}
	$\lim_{N \rightarrow \infty} \theta_N^M \approx \theta_2^M \times [] e ( N-1)^{-1} ]$
	%\end{align}
	where $e$ is Euler's number. Figure \ref{magic}(d) shows results for magic angle combinations in stacks of four layers, defined by three twist angles, and three integers, $l , m , n$. Calculations for the chiral helical case, $\theta_{1} = \theta_{23} = \theta_{34} = \theta_0$, can be found in~\cite{SI}.  
	
	The magic angles where flat bands appear depend on the integers $m, n$, and on the displacement $\mathbf{D}$. To a first approximation, we can approximate the electronic density of states of the system by an average over the values of $\mathbf{D}$. This approximation can be expected to become exact in the limit $\ell_{m2} / \ell_m \sim 1 / \theta_0 \rightarrow \infty$, and also when the dependence of the band structure on $\mathbf{D}$ is small. 
	
	The flat bands at the magic angles reported here show a greater degree of plasmon screening compared to tBG and thus, a greater resilience against distortion from interaction effects. They also have a finite Berry curvature and, typically, non zero Chern numbers, see~\cite{SI}.

	{\it Results. Analysis of the trilayer studied in~\cite{Uetal23}.}
	We now analyze the experimental situation reported in~\cite{Uetal23}. The trilayer studied there shows angles $\theta_{12} \approx 1.42^\circ, \theta_{23} \approx - 1.88^\circ$, which are well approximated by $\theta_{12} = 3 \theta_0$ and $\theta_{23} = -4 \theta_0$ with $\theta_0 \approx 0.47^\circ$. Results for local and average densities of states are shown in Fig. \ref{fig:dos34} (we use $\{ t_{AA} , t_{AB} \} = \{ 0.0797 , 0.0975 \}$ eV, and $\gamma_0 = 3.06$ eV, which implies $v_F \approx 987$ Km/s, in order to compare with~\cite{Uetal23}). As shown in the figure, the dependence of the density of states on the displacement $\mathbf{D}$ is very weak, which suggests that the weighted average shown in Fig. \ref{fig:dos34} (d) gives a good approximation to the exact value.

	\begin{figure}[htbp]
		\centering
		\includegraphics[width=0.5\textwidth]{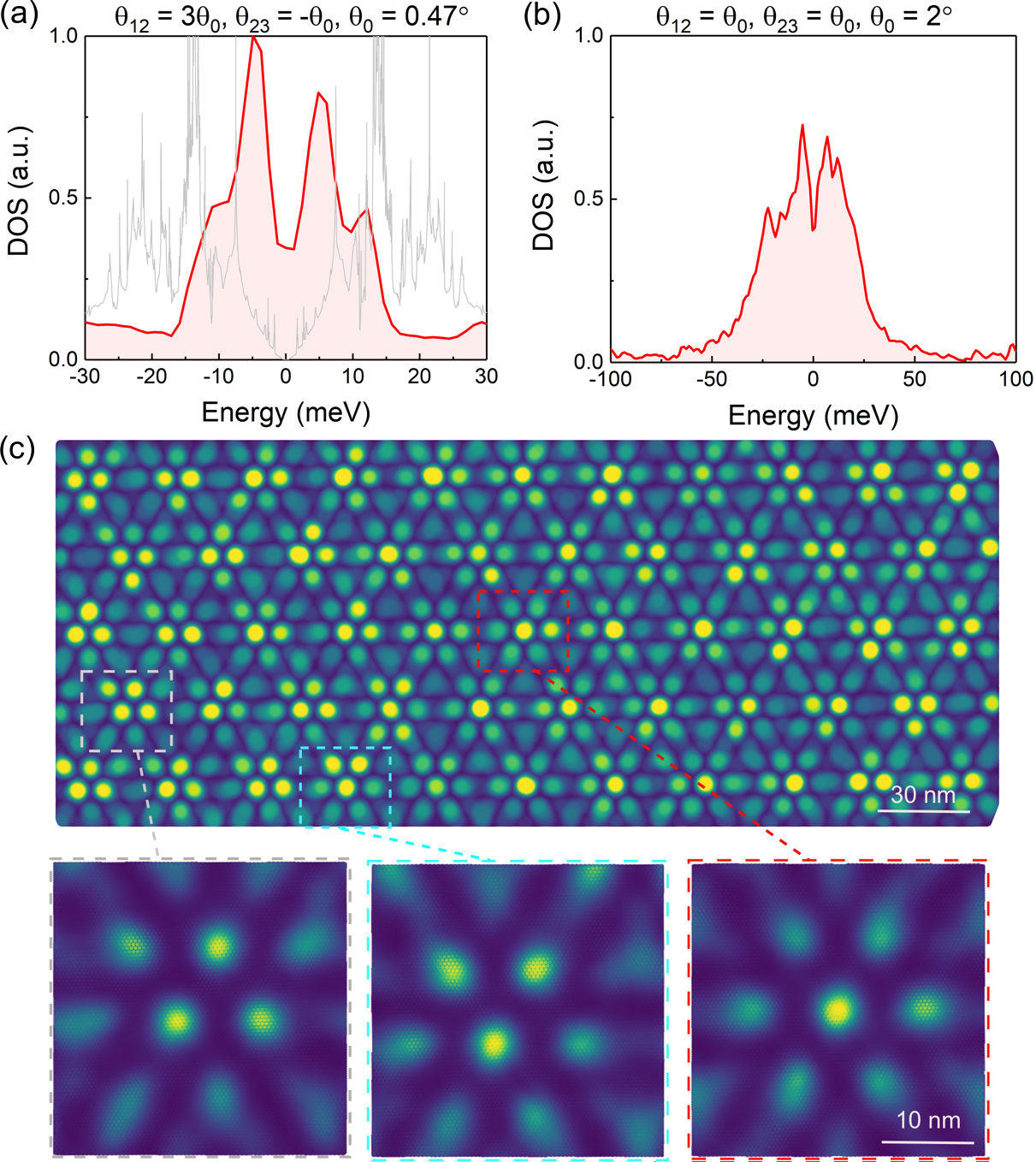}
		\caption{The tight-binding results. (a) The density of states for a trilayer with angles $\theta_{12} = 1.41^\circ$, $\theta_{23} = - 1.88^\circ$. The gray line is the result in the chiral limit. (b) The density of states for a trilayer with angles $\theta_{12} = 2^\circ$, $\theta_{23} = 2^\circ$. (c) The local density of states at energy of 7 meV in the real space for trilayer in (a), with a zoom in of the states with displacements $\mathbf{D}$ marked by dashed square with colors. Red and gray correspond to the color in Fig. \ref{magic}(a), and blue are in the middle region.}
		\label{fig:tb}
	\end{figure}
	
	Figure \ref{fig:tb}(a) is the tight-binding (TB) density of states for trilayer with $\theta_{12} \approx 1.42^\circ, \theta_{23} \approx - 1.88^\circ$, which is obtained via the parameters $\gamma_0=3.1$ eV and $\gamma_1=0.43$ eV~\cite{li2023tbplas,yu2019dodecagonal,SK1954simplified}, the same as the ones in~\cite{Uetal23}. The discrepancy between the TB and chiral limit results is due to the difference of the Fermi velocity. Moreover, if we reduce the Fermi velocity, a flat band appears in the charge neutrality point (not shown here). The states of the peak at the energy of 7 meV in Fig. \ref{fig:tb}(a) is highly localized with an independence of the displacement $\mathbf{D}$, shown in Fig. \ref{fig:tb}(c), which is in agreement with the results in Fig. \ref{fig:dos34} in the chiral limit. 
	
	We extend our discussion to another configuration, the helical trilayer graphene with $\theta_{12}=\theta_{23}=\theta_0$. One magic angle is $\theta_0=2^\circ$~\cite{MZY23}, which has peaks located at the charge neutrality point, shown in Fig. \ref{fig:tb}. If we consider the lattice relaxation~\cite{thompson2022lammps,brenner2002second,kolmogorov2005registry}, the magic angle reduces to $1.57^\circ$~\cite{Detal23}. For $AAA$ helical trilayer graphene with identical twist angles, the moir\'e-of-moir\'e structure locally relaxes into large areas that contains only one moir\'e pattern. That is, a large region of the periodic $ABA$ could be reconstructed if a strong intralayer and interlayer potential is utilized. The states of the flat band in the relaxed system is highly dependent on the displacement $\mathbf{D}$ (see~\cite{SI} Fig. 10), which is consistent with the results in Fig. \ref{magic}(a). Furthermore, the helical trilayer graphene with $\theta_{12}=\theta_{23}$ could be an ideal platform for realization of the chiral limit due to significant corrugation, or via heterostrain engineering~\cite{Detal23}.
	
	{\it Conclusions.}
	%We have analyzed the existence of flat bands in twisted stacks of graphene layers where the twist angles do not allow for the definition of a simple moir\'e structure. We have generalized previous works where approximate moir\'e structures in trilayers~\cite{CWG19,MRB19,WZS20,MGM23,PT23,GMM23,Detal23}. By analyzing the chiral limit~\cite{SGG12,TKV19} we show that flat bands appear for all combinations of twist angles where the angles are close to coprime numbers. We have shown that a local approximation in terms of two different periodicities, as developed in~\cite{MGM23}, describes well the electronic structure of twisted multilayers. These results are consistent with large scale tight binding calculations~\cite{MZY23}. The analysis presented here opens the way to the study of topological features in many types of twisted multilayers, and they suggest that the rich phase diagram found in twisted bilayer graphene may exist in many other graphene stacks.
	We have considered the existence of flat bands in twisted stacks of graphene layers where the twist angles do not allow for the definition of a simple \moire structure, generalizing previous works~\cite{CWG19,MRB19,WZS20,MGM23,PT23,GMM23,Detal23,NKK23} which have focused on special configurations yielding exact or approximate \moire structures in trilayers. By analyzing the chiral limit~\cite{SGG12,TKV19} we conclusively demonstrate that flat bands appear for all combinations of twist angles where the angles are close to multiples of coprime integers. 
	%These magic trilayers, often with both twist angles larger than the tBG magic angle, should be more resistant to relaxation effects, and are expected to be more robust against Hartree effects than those of tBG due to enhanced plasmon screening~\cite{SI}, making our calculations more applicable to real experiments. 
	Our further extension to generic, incommensurate angles~\cite{SI}  connects the previously disparate magic points of specific trilayer configurations into an extended magic line. 
	Building on the local approximation as developed in~\cite{MGM23} for the specific case of helical trilayers with nearly equal twist angles, we accurately describe the electronic structure of generic twisted multilayers even away from the magic angle, with our results consistent with both large scale tight binding calculations~\cite{MZY23} and contemporary experiments~\cite{Uetal23}. The analysis presented here opens the way to the study of topological features in many types of twisted multilayers, and they suggest that the rich phase diagram found in twisted bilayer graphene may exist in many other graphene stacks.
	
	{\it Note added.}
	We have noticed ref.(\cite{PT23b}, posted a day before our paper. This reference discusses topics related to our work. Insofar the two manuscript overlap, the results are in agreement.
	
	{\it Acknowledgements.}
	We are thankful to B. Amorim, B. A. Bernevig, and E. Castro for helpful discussions.
	IMDEA Nanociencia acknowledges support from the ``Severo Ochoa" Programme for Centres of Excellence in R\&D (CEX2020-001039-S / AEI / 10.13039/501100011033).
	F.G. acknowledges funding from the European Commission, within the Graphene Flagship, Core 3, grant number 881603 and from grants NMAT2D (Comunidad de Madrid, Spain), SprQuMat (Ministerio de Ciencia e Innovaci\'on, Spain) and financial support through the (MAD2D-CM)-MRR MATERIALES AVANZADOS-IMDEA-NC.  DCWF, MA, LP and  SA acknowledge support from the Singapore National Research Foundation Investigator Award (NRF-NRFI06-2020-0003).  Z.Z. acknowledges support funding from the European Union's Horizon 2020 research and innovation programme under the Marie Skłodowska-Curie grant agreement No 101034431. Numerical calculations presented in this paper have been performed on the supercomputing system in the Supercomputing Center of Wuhan University.

	%%%%%%%%%% Merge with supplemental materials %%%%%%%%%%
	\newpage
	
	\pagebreak
	\widetext
	%\onecolumn
	\begin{center}
		\textbf{\large Supplemental Material for:\\Extended magic phase in generalized trilayer graphene}
	\end{center}
	%%%%%%%%%% Merge with supplemental materials %%%%%%%%%%
	%%%%%%%%%% Prefix a "S" to all equations, figures, tables and reset the counter %%%%%%%%%%
	\setcounter{equation}{0}
	\setcounter{figure}{0}
	\setcounter{table}{0}
	\setcounter{page}{1}
	\makeatletter
	\renewcommand{\theequation}{S\arabic{equation}}
	\renewcommand{\thefigure}{S\arabic{figure}}
	%\renewcommand{\bibnumfmt}[1]{[S#1]}
	%\renewcommand{\citenumfont}[1]{S#1}
	%%%%%%%%%% Prefix a "S" to all equations, figures, tables and reset the counter %%%%%%%%%%

	%%%%%%%%%%%%%%%%%%%%%%%%%%%%%%%%%%%%%%%%%%%%%
	\section{Definition of periodicities.}
	For completeness, we specify the two relevant periodicities in a twisted graphene trilayer with almost commensurate twist angles, $\theta_{12}$ and $\theta_{23}$ following the steps described in~\cite{MGM23}. We write the two angles as:
	\begin{align}
		\theta_{12} &= m \theta_0 + m \delta \theta\nonumber \\
		\theta_{23} &= n \theta_0 - n \delta
		\theta
	\end{align}
	where $m , n$ are coprime integers, and we assume that $\theta_0 \ll 1$ and $\delta \theta \ll \theta_0$. The angles $\theta_{12}$ and $\theta_{23}$ define moir\'e structures. The vectors connecting the points $\vec{K}$ and $\vec{K}'$ of the respective Brillouin zones are:
	\begin{align}
		\Delta \vec{K}_{12} &\approx \frac{4 \pi}{3 d} \left\{ - \frac{m^2 ( \theta_0^2 + 2 \theta_0 \delta \theta )}{2} , m \theta_0 + m \delta \theta \right\} \approx m \vec{g}_0 + m \delta \vec{g}
		\nonumber \\
		\Delta \vec{K}_{23} &\approx \frac{4 \pi}{3 d} \left\{ - \frac{n^2 ( \theta_0^2 - 2 \theta_0 \delta \theta )}{2} , n \theta_0 - n \theta_0 \right\} \approx n \vec{g}_0 - n  \delta \vec{g}
	\end{align}
	where we have assumed that $\theta_0 \ll 1$.
	We obtain:
	\begin{align}
		\vec{g}_0 &\approx \frac{4 \pi}{3 d} \left\{ \frac{m+n}{4} \theta_0^2 - \frac{m-n}{2} \theta_0 \delta \theta , \theta_0 \right\} \approx \frac{4 \pi}{3 d} \left\{ 0, \theta_0 \right\}
		\nonumber \\
		\delta \vec{g} &\approx \frac{4 \pi}{3 d} \left\{ - \frac{m-n}{4} \theta_0^2 - \frac{m+n}{2} \theta_0 \delta \theta , \delta \theta \right\}
		\nonumber \\
	\end{align}
	Vectors $\vec{g}_0$ and $\delta \vec{g}$ and three equivalent vectors rotated by $\pm 120^\circ$ define the relevant periodicities of the trilayer. The two vectors satisfy $| \delta \vec{g} | \ll | \vec{g_0} |$, so that, in real space, the periodicity associated with $\delta \vec{g}$ changes much more slowly than that asociated to $\vec{g}_0$. Locally, $\vec{g}_0$ defines a moir\'e lattice with unit vector $\ell_m \approx d / \theta_0$, where $d$ is the lattice constant of graphene.
	
	In the following, for simplicity, we assume that $\delta \theta \ll \theta_0^2$, although this assumption is not crucial. Then, $\delta \vec{g}$ defines a new periodicity, with unit cell rotated $90^\circ$ with respect to the unit cell defined by $\vec{g}_0$ and a lattice unit of length $\ell_{m2} \approx ( 4 d ) / ( | m - n | \theta_0^2 ) \gg \ell_m$. Note, finally, that our definition of $\delta \vec{g}$, for $m \ne 1$ or $n \ne 1$, differs from the one used in~\cite{MGM23}, although both definitions scale as $| \delta \vec{g} | \propto \theta_0^2$.
	
	The analysis described above can be extended to an arbitrary number, $N$, of layers, and $N_1$ twist angles between nearest layers, provided that these angles are almost commensurate, defined by $N-1$ coprime integers. Such a calculation leads to a periodicity, $\vec{g}_0$, defined by an angle $\theta_0$, and $N-2$ additional periodicities defined by vectors $\delta \vec{g}_{i = 1 , \cdots , N-2}$ such that $| \delta \vec{g}_i | \ll | \vec{g}_0 |$. Insofar as the role of $\delta \vec{g}$ can be reduced to an overall displacement between the layers (see below) the two definitions should give the same results.
	
	%%%%%%%%%%%%%%%%%%%%%%%%%%%%%%%%%%%%%%%%%%%%%%%%%%%%%
	\section{Slow periodicity as a displacement between layers.}
	
	%We analyze a continuum approximation to the electronic structure of a graphene trilayer defined by the angles $\theta_{12}$ and $\theta_{23}$, where 2 labels the central layer, and 1 and 3 the outer layers. We assume that the angles $\theta_{12}$ and $\theta_{23}$ are commensurate, so that $\theta_{12} = m \theta_0$ and $\theta_{23} = n \theta_0$, where $m$ and $n$ are integers. We define a moir\'e lattice based on the angle $\theta_0$. The length of the moir\'e lattice vector is $\ell_M = d / [2 \sin ( \theta_0 / 2) ]$ where $d$ is the length of the graphene lattice unit vector. We assume that the top and bottom layers are rotated with respect to the central layer using an axis which intersects the trilayer at a point with an $AAA$ arrangement, where $A$ is a sublattice label. This assumption will be relaxed later.
	
	We now elaborate on the argument that the slow periodicity, $\delta \vec{g}$, identified in the previous section can be interpreted as a local displacement of the top layer with respect to the bottom layer~\cite{MGM23}.
	
	\begin{figure}[h!]
		\includegraphics[width=2.5in]{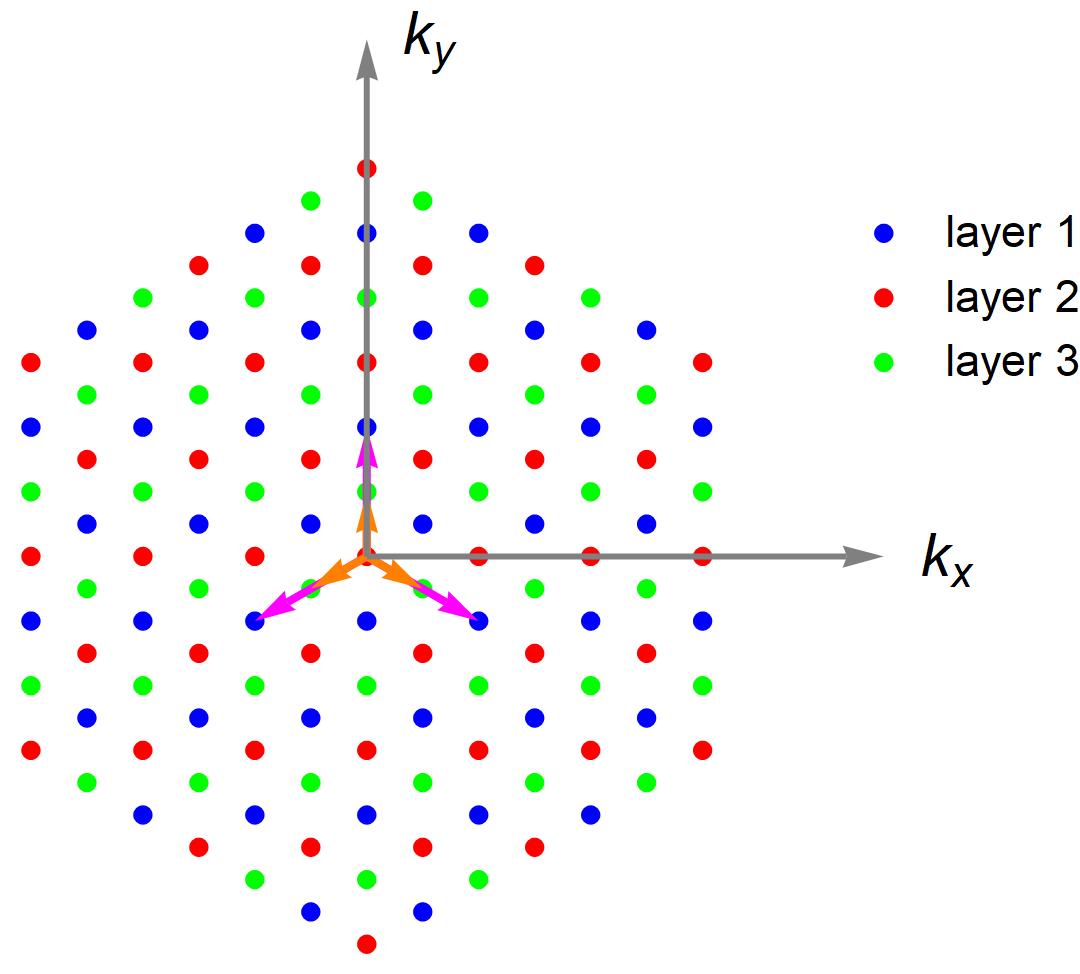} 
		\caption{Sketch of the momenta of the Bloch waves included in the continuum model, see eq.(\ref{hamilt}), for $m=2, n= 1$. The magenta arrows represent the vectors $m \vec{g}_i$, and the orange arrows represent the vectors $n \vec{g}_i$, see eq.(\ref{vectors}).}
		\label{fig:bloch}
	\end{figure}
	
	The interlayer tunneling part of the continuum Hamiltonian \cite{LPN07,BM11}, in real space, is:
	\begin{align}
		{\cal H}_T &= {\cal H}_T^{12} + {\cal H}_T^{23}
		\nonumber \\
		{\cal H}_T^{12} &= \sum_{j=1,2,3} T_j e^{i m \vec{g}_j \cdot \vec{r}}
		\nonumber \\
		{\cal H}_T^{23} &= \sum_{j=1,2,3} T_j e^{i n \vec{g}_j \cdot \vec{r}}
		\label{hamilt}
	\end{align}
	where $T_j$ define inter sublattice $2 \times 2$ matrices:
	\begin{align}
		T_j &\equiv \left( \begin{array}{cc} t_{AA} &t_{AB} e^{i \frac{2 \pi ( j- 1)}{3}} \\t_{AB} e^{-i \frac{2 \pi ( j- 1)}{3}} &t_{AA} \end{array} \right)
		\label{matrices}
	\end{align}
	and:
	\begin{align}
		\vec{g}_1 &= \frac{\ell_m}{\sqrt{3}} \left\{ 0 , 1 \right\}
		\nonumber \\
		\vec{g}_2 &= \frac{\ell_m}{\sqrt{3}} \left\{ \frac{\sqrt{3}}{2} , - \frac{1}{2} \right\}
		\nonumber \\
		\vec{g}_3 &= \frac{\ell_m}{\sqrt{3}} \left\{ - \frac{\sqrt{3}}{2} , - \frac{1}{2} \right\}
		\label{vectors}
	\end{align}
	
	The momenta of the Bloch waves used in the calculation are shown in Fig. \ref{fig:bloch}. 
	
	We can generalize the tunneling Hamiltonian to the case where the \moire pattern of layers 1 and 2 is displaced by a shift $\vec{\mathbf{D}}_1$ and the \moire pattern between layers 2 and 3 is displaced by a shift $\vec{\mathbf{D}}_3$~\cite{LWM19}. In the initial Hamiltonian, the rotation is around a point with $AA$ stacking between layers 1 and 2, and $AA$ stacking between layers 2 and 3, that is, $AAA$ stacking. The shifts lead to the Hamiltonian:
	%We can generalize the tunneling Hamiltonian, and assume that layer 1 is displaced by a shift $\vec{\mathbf{D}}_1$ and layer 3 is displaced by  a shift $\vec{\mathbf{D}}_3$~\cite{LWM19}. In the initial hamiltonian, the rotatio is around a point with $AA$ stacking between layers 1 and 2, and $AA$ stacking between layers 2 and 3, that is, $AAA$ stacking. The shifts lead to the hamiltonian:
	\begin{align}
		{\cal H}_T &= {\cal H}_T^{12} + {\cal H}_T^{23}
		\nonumber \\
		{\cal H}_T^{12} &= \sum_{j=1,2,3} T_j e^{i m \vec{g}_j \cdot \left( \vec{r} - \vec{\mathbf{D}}_1 \right)}
		\nonumber \\
		{\cal H}_T^{23} &= \sum_{j=1,2,3} T_j e^{i n \vec{g}_j \cdot \left( \vec{r} - \vec{\mathbf{D}}_3 \right)}
		\label{hamiltd}
	\end{align}
	Each Bloch wave in the calculation, see Fig. \ref{fig:bloch}, can be modified by a gauge factor, $e^{i \vec{G}  \cdot \vec{\mathbf{D}^\prime} }$, where $\vec{G}$ is the momentum of the wave. This gauge transformation induces a phase in the interlayer tunnelings, which is equivalent to a redefinition of the shifts:
	\begin{align}
		\vec{\mathbf{D}}_1' &= \vec{\mathbf{D}}_1 - \vec{\mathbf{D}}^\prime
		\nonumber \\
		\vec{\mathbf{D}}_3' &= \vec{\mathbf{D}}_3 - \vec{\mathbf{D}}^\prime
	\end{align}
	This result is independent of the integers $m$ and $n$, and of the angle $\theta_0$. An equal shift, $\vec{\mathbf{D}}_1 = \vec{\mathbf{D}}_3$, of the two \moire patterns can be gauged away. This implies that the calculated electronic structure is only dependent on the value of $\vec{\mathbf{D}}_1 - \vec{\mathbf{D}}_3=\mathbf{D}$. 
	
	%We now analyze the trilayer as a combination of two different periodicities, as in~\cite{MGM23}. We assume that $\theta_{12} = m \theta_0$ and $\theta_{23} = n \theta_0$, and define:
	%\begin{align}
	%   \frac{4 \pi}{3 d} \{ 1 - \cos ( m \theta_0 ) , \sin ( m \theta_0 ) \} &\approx  \frac{4 \pi}{3 d} \left\{ \frac{m^2 \theta_{0}^2}{2} , m \theta_{0} \right\} =  
	%    \nonumber \\ &= m \vec{g}_1 + m \delta \vec{g}_1
	%   \nonumber \\
	%    \frac{4 \pi}{3 d} \{ 1 - \cos ( m \theta_0 ) , \sin ( n \theta_0 ) \} &\approx  \frac{4 \pi}{3 d} \left\{ \frac{n^2 \theta_{0}^2}{2} , n \theta_{n} \right\} =  
	%  \nonumber \\ &= m \vec{g}_1 - n \delta \vec{g}_1
	%    \label{period}
	%\end{align}
	%The equations for $\vec{g}_ 2 , \vec{g}_3 , \delta \vec{g}_2 , \delta \vec{g}_3$ can be obtained by rotating eq.(\ref{period}). The solutions of these equation are:
	%\begin{align}
	%  \vec{g}_1 &\approx \frac{4 \pi}{3 d} \left\{ \frac{(m+n) \theta_0^2}{2} , \theta_0 \right\} \approx \frac{4 \pi}{3 d} \{ 0 , \theta_0 \}
	%   \nonumber \\
	%    \delta \vec{g}_1 &\approx \frac{4 \pi}{3 d} \left\{ \frac{(m-n) \theta_0^2}{2} , 0 \right\}
	%\end{align}
	%where we have expanded each term to lowest order in $\theta_0$.
	
	\begin{figure}[h!]
		%\begin{tabular}{lr}
		\includegraphics[width=0.6\textwidth]{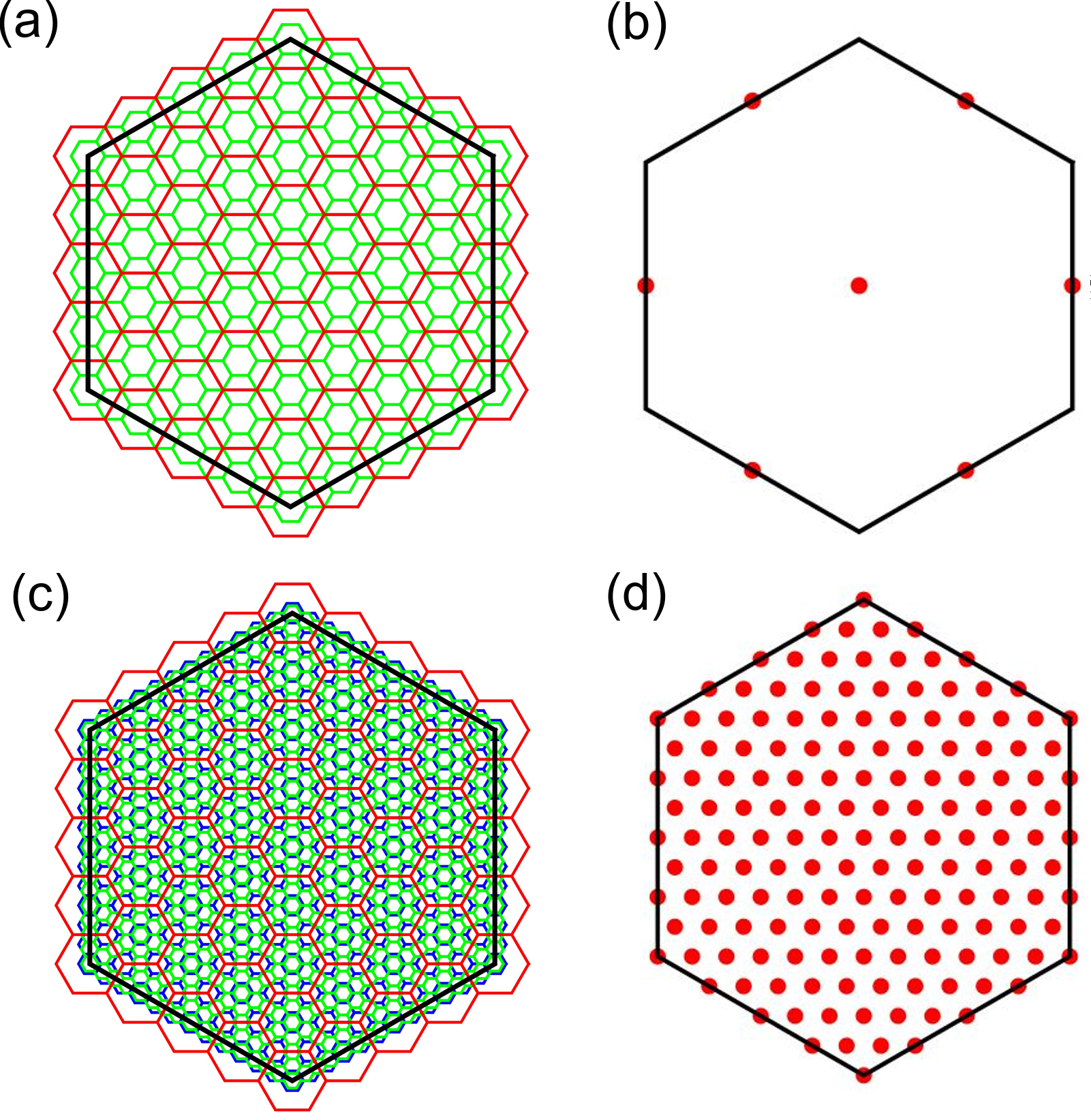} 
		%\includegraphics[width=2.5in]{trilayer_uc_points_sketch_12.jpg} 
		%\\
		%\includegraphics[width=2.5in]{trilayer_uc_sketch_34.jpg} &
		%\includegraphics[width=2.3in]{trilayer_uc_points_sketch_34.jpg} 
		%\end{tabular}
		\caption{Sketch of the real space unit cell structure for (a) trilayer with $\{ m , n \} = \{ 1 , 2 \}$ and (c) trilayer with $\{ m , n \} = \{ 3 , -4 \}$. The black hexagon defines the large unit cell, associated with moir\'e length $\ell_{m2}$. The red  hexagons show the moir\'e structure associated with angle $\theta_0$ and moir\'e length $\ell_m$. Blue and green hexagons show moir\'es associated with angles $\theta_{12}$ and $\theta_{23}$. Note that the moir\'e lattices associated to $\theta_0$ and $\theta_{12}$ coincide. Special points in the momentum space unit cell with the $AAA$ electronic structure for (b) $\{ m , n \} = \{ 1 , 2 \}$ and (d) $\{ m , n \} = \{ 3 , -4 \}$ cases.}
		\label{fig:unit_cell}
	\end{figure}
	
	The two periodicities defined by $\vec{g}_0$ and $\delta \vec{g}$ in the previous section modify the continuum Hamiltonian, Eq. (\ref{hamilt}), leading to:
	\begin{align}
		{\cal H}_T^{12} &= \sum_{j=1,2,3} T_j e^{i m  \vec{g}_{0j}  \cdot   \vec{r}  }
		e^{i m   \delta \vec{g}_j  \cdot  \vec{r}  }
		\nonumber \\
		{\cal H}_T^{23} &= \sum_{j=1,2,3} T_j e^{i n  \vec{g}_{0j}  \cdot   \vec{r}  }
		e^{- i n   \delta \vec{g}_j  \cdot  \vec{r}  }
		\label{hamilt3}
	\end{align}
	The vectors $\delta \vec{g}_j$ define a periodicity with unit vectors in real space:
	\begin{align}
		\{ \vec{D}_1 , \vec{D}_3 \}   &= \frac{2 \pi}{3 d | \delta \vec{g}_i  |}      \{ \vec{R}_1 , \vec{R}_2 \} \approx \frac{1}{\theta_0^2}  \{ \vec{R}_1 , \vec{R}_2 \}
	\end{align}
	where $\vec{R}_1$ and $\vec{R}_2$ are the lattice vectors of an untwisted graphene layer.
	
	We now assume that the arguments $\delta \vec{g}_j \cdot \vec{r}$ vary slowly with respect the functions $\vec{g}_{0j} \vec{r}$, and assume that the value of $\vec{r} = \vec{\mathbf{D}}$ in $\delta \vec{g}_j \cdot \vec{r}$ is constant. Then, using Eq. (\ref{hamiltd}) we obtain that the local Hamiltonian is equivalent to displacements of the top and bottom layers by $\pm ( d / \ell_{m2} ) R_{\pi/2} (\vec{\mathbf{D}} )$, where $R_\theta$ is a rotation operator.

	For $m \ne 1$ or $n \ne 1$ the loops in momentum space which connect a momentum in the central layer to itself through a path which visits both the top and bottom layers require at least $m \times n$ steps. The increased periodicity of these loops for $m \ne 1$ or $n \ne 1$ imply that the band structure in real space repeats itself for a displacement periodicity defined by the unit vectors:
	\begin{align}
		\{ \tilde{\vec{D}}_1 , \tilde{\vec{D}}_3 \} &= \frac{1}{mn} \{\vec{D}_1 , \vec{D}_3 \}
	\end{align}
	This result implies that the actual periodicity in real space is given by $\ell_{m2} / ( m n )$. A sketch of the different regions in the unit cell, and the  points where the band structure repeats itself is shown in Fig. \ref{fig:unit_cell}.
	
	\begin{figure}[t]
		%\centering
		%\begin{tabular}{cc}
		\includegraphics[width=0.95\linewidth]{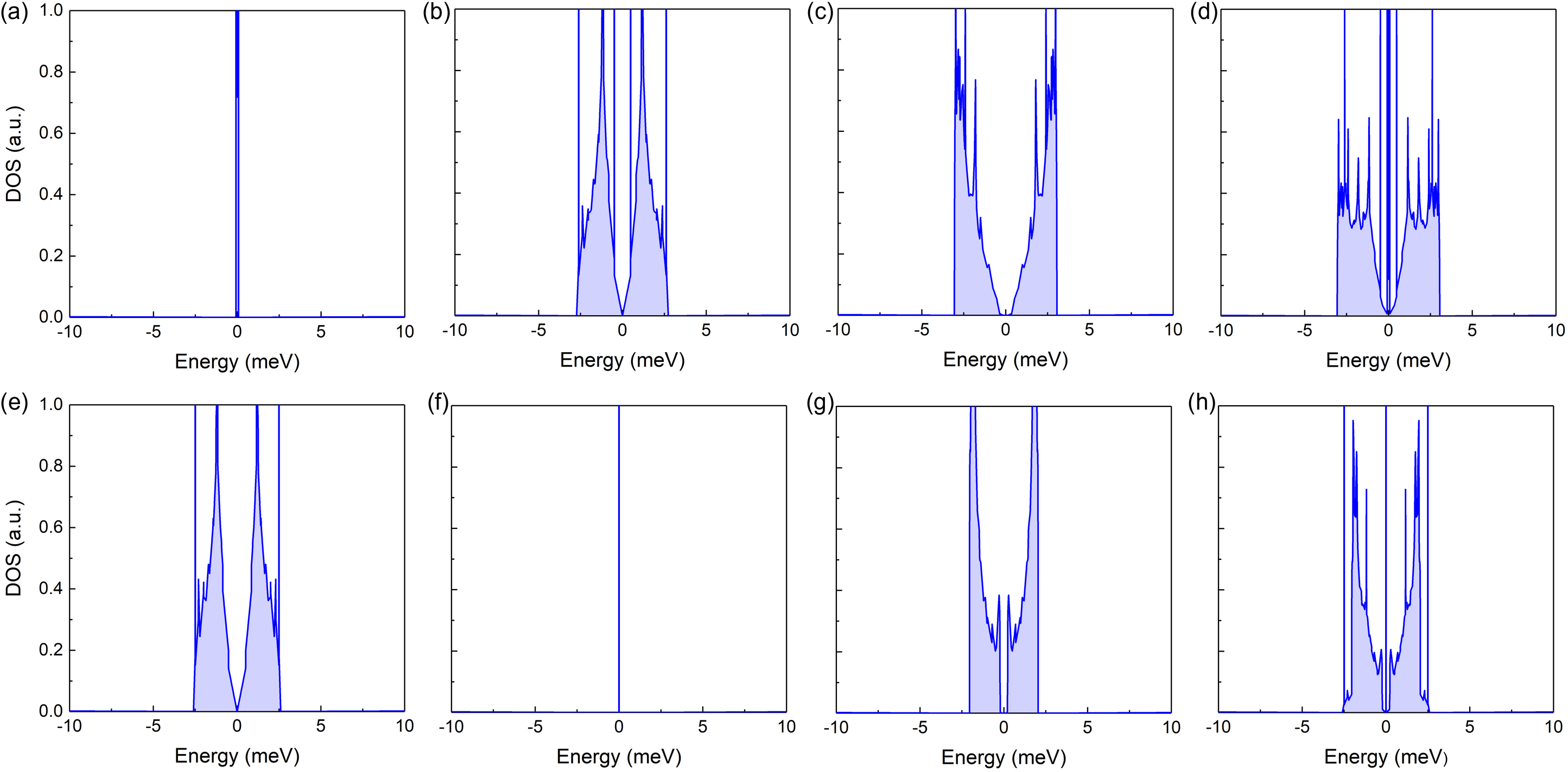} 
		%\includegraphics[width=0.40\linewidth]{dos2m1k.jpg} \\
		%\includegraphics[width=0.20\linewidth]{dos2m1kt.jpg} &
		%\includegraphics[width=0.20\linewidth]{dos2m1kt.jpg}
		%\end{tabular}
		\caption{Results for the magic angles in the chiral limit and $\theta_{12}=2 \theta_0 , \theta_{23} = \theta_0$. (a)
			- (d): The local DOS for a magic angle $\theta_0 = 1.163^\circ$ with three displacements and their average, as in Fig. 3 of the main text. (e) - (h): As in the top panel, but for a magic angle $\theta_0 = 1.14^\circ$. }
		\label{fig:dos2m1}
	\end{figure}
	
	\begin{figure}[t]
		%\centering
		%\begin{tabular}{cc}
		\includegraphics[width=0.95\linewidth]{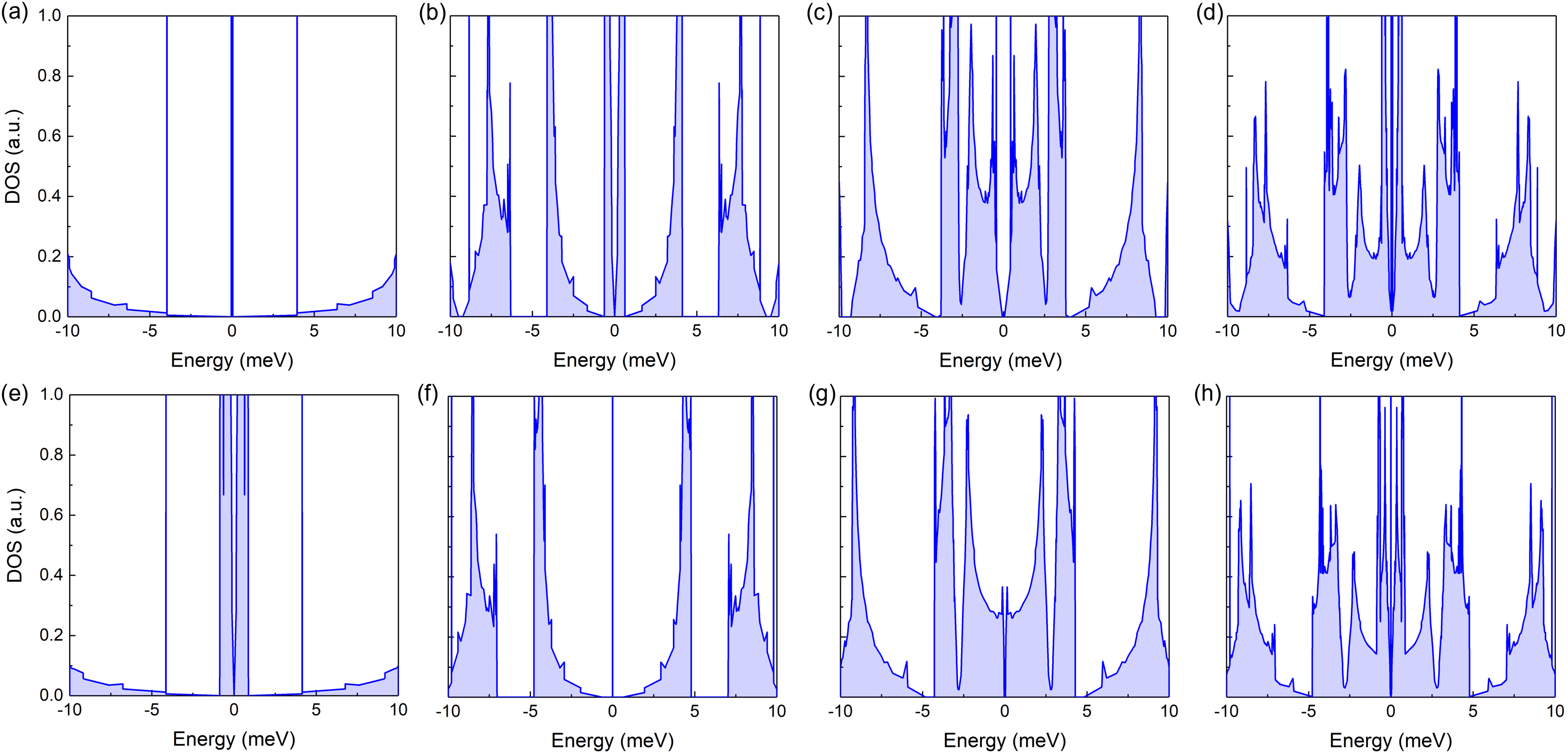} 
		%\includegraphics[width=0.40\linewidth]{dos3m2k.jpg} \\
		%\includegraphics[width=0.20\linewidth]{dos3m2kt.jpg} &
		%\includegraphics[width=0.20\linewidth]{dos3m2kt.jpg}
		%\end{tabular}
		\caption{As in Fig. \ref{fig:dos2m1}  but for  $\theta_{12}=3 \theta_0 , \theta_{23} = 2 \theta_0$. (a) - (d): The local DOS for a magic angle $\theta_0 = 0.6^\circ$ with three displacements and their average as in Fig. 3 of the main text. (e) - (h): Results for a magic angle $\theta_0 = 0.59^\circ$.}
		\label{fig:dos3m2}
	\end{figure}
	
	Densities of states (DOS) for two choices of the magic angle in the chiral limit, and $\{ m , n \} = \{ 2 , 1 \}$ are shown in Fig. \ref{fig:dos2m1}. Similar results for $\{ m , n \} = \{ 3 , 2 \}$ are shown in Fig. \ref{fig:dos3m2}.
	%%%%%%%%%%%%%%%%%%%%%%%%%%%%%%%%%%%%%%%%%%%%%%%%%%%%%
	\section{Extension to multilayers with $N>3$.}
\begin{figure}[t!]
		\includegraphics[width=\textwidth]{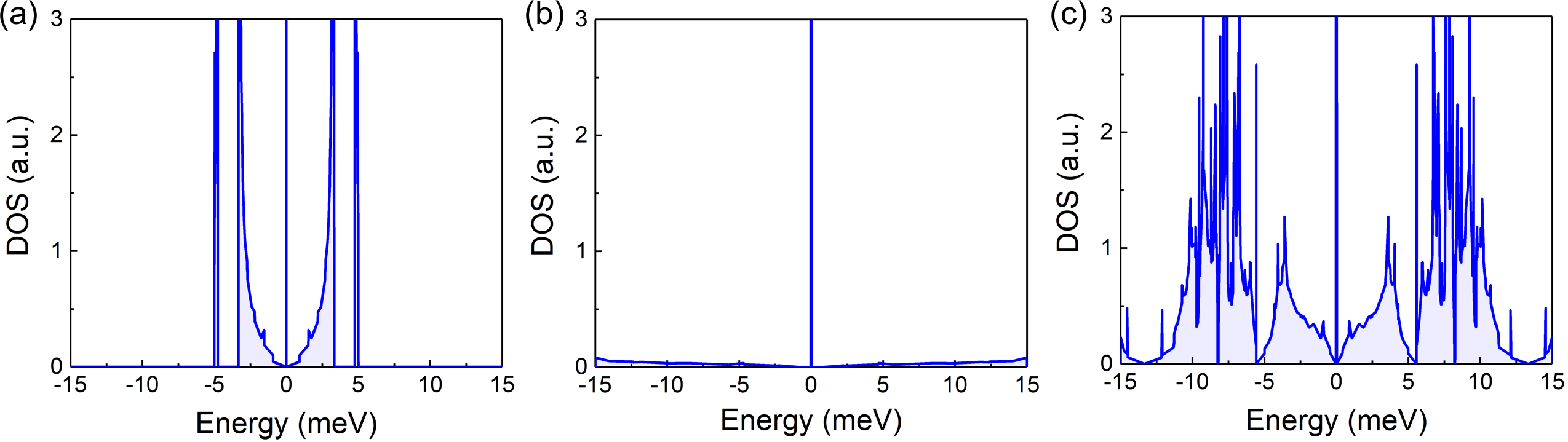} 
		\caption{DOS for magic angles in the chiral limit of four layer stacks with $\theta_{12} = \theta_{23} = \theta_{34} = \theta_0$ (helical arrangement). (a) $AAAA$ stacking, $\theta_0 = 0.55^\circ$. (b) $AABB$ stacking, $\theta_0 = 0.395^\circ$. (c) $ABAB$ stacking, $\theta_0 = 0.299^\circ$.}
		\label{fig:four}
	\end{figure}

	The analysis discussed previously can be extended to multilayers with any number of layers greater than three. Results for the density of states in the chiral limit of an helical stack of four layers ($\theta_{12} = \theta_{23} = \theta_{34} = \theta_0$) at different magic angles are shown in Fig. \ref{fig:four}.
	
	It is also instructive to generalize the alternative method  discussed in~\cite{PT23}. The analysis addresses the magic angles, in the chiral limit, of a stack of $N$ layers, where each layer is twisted by the same angle, $\theta$, and in the same direction,  with respect to the previous layer. 
	
	The magic angles for $N$ layers are determined by an $N-$component spinor which is the solution of $N$ first order differential equations, similar to the case of a 2-component spinor, $\{ \psi_{\vec{k}} , \bar{\psi}_{\vec{k}} \}$, which determines the magic angles in twisted bilayer graphene. The solution found in~\cite{PT23} describes the three component spinor for the trilayer in terms of three combinations of products of combinations of these wavefunctions:
	\begin{align}
		\Psi_3 &\equiv \left\{ \begin{array}{c} 
			\psi_{\vec{k}_1}  \psi_{\vec{k}_2} \\
			\frac{1}{\sqrt{2}} \left( \psi_{\vec{k}_1}  \bar{\psi}_{\vec{k}_2} + \bar{\psi}_{\vec{k}_1} \psi_{\vec{k}_2} \right) \\
			\bar{\psi}_{\vec{k}_1}  \bar{\psi}_{\vec{k}_2}
		\end{array}
		\right.
	\end{align}
	A similar equation can be written for an $N-$component spinor:
	\begin{align}
		\Psi_3 &\equiv \left\{ \begin{array}{c} 
			\prod_{i=1}^{i=N} \psi_{\vec{k}_i}   \\
			\frac{1}{\sqrt{N}} 
			\sum_{j=1}^{j \le N} \bar{\psi}_{\vec{k}_j} \prod_{i=1,i\ne j}^{i=N} \psi_{\vec{k}_i}
			\\
			\frac{1}{\sqrt{N ( N-1)}}  \sum_{j=1,k=1,k\ne j}^{j=N,k=N} \bar{\psi}_{\vec{k}_j} \bar{\psi}_{\vec{k}_k} \prod_{i=1,i\ne j,i \ne k}^{i=N} \psi_{\vec{k}_i}
			\\
			\cdots
			\\
			\prod_{i=1}^{i=N} \bar{\psi}_{\vec{k}_i}
		\end{array}
		\right.
		\label{ml_magic}
	\end{align}
	We defined $\alpha_2 =  [ v_F ( 4 \pi ) / ( 2 d \sin  \theta_M / 2 ) ] / t_{AB}$, where $\theta_M$   is the magical angle for a twisted bilayer. Then, Eq. (\ref{ml_magic}) has a solution for:
	\begin{align}
		\alpha_N &= \frac{4 \pi v_F }{d t_{AB}} \left[ 2  \sin \left( \frac{\theta_M^N}{2} \right) \right] = \alpha_2 \left( \frac{1}{(N-1) !} \right)^{\frac{1}{N-1}}
	\end{align}
	We have checked that this equation gives the correct solution for $\alpha_3 = \alpha_2 / \sqrt{2}$ and $\alpha_4 = \alpha_2 \times 1 / \sqrt[3]{6}$  agrees with our numerical results for a stack of four layers. For $N \rightarrow \infty$ we find:
	\begin{align}
		\lim_{N \rightarrow \infty} \theta_M^N &\approx \frac{d t_{AB}}{4 \pi v_F} \frac{e}{N-1}
	\end{align}
	where $e$ is Euler's number. This result implies that a small shear force applied to the top and bottom layers of a graphite stack can lead to a peak at the density of states near the Fermi level.

	%%%%%%%%%%%%%%%%%%%%%%%%%%%%%%%%%%%%%%%%%%%%%%%%%%%%%%
	\section{Validity of the local Hamiltonian expansion}
	%\begin{figure}[h]
	%	\centering
	%   \includegraphics[width=0.7\linewidth]{moml_diag.pdf}
	%	\includegraphics[width=0.7\linewidth]{moml.pdf}
	%	\caption{Top: Sketch of connected momentum states between layers with misaligned \moire patterns (orange and purple), resulting in a \moire-of-\moire modulation. The lengthscale of this modulation is determined by the closest approach of second layer lattice sites formed from the two \moire patterns, indicated with a dotted line. The local expansion used in numerical calculation neglects this misalignment and, for commensurate angles, collapses this distance to zero. Bottom: Log-scale plot of the \moire-of-\moire length scale, $l_{m2}$, in microns, as a function of twist angles. The left diagonal in white marks the line $\theta_{12}=-\theta_{23}$, where there is no misalignment and $l_{m2}$ diverges. $l_{m2}$ also diverges when either angle is zero and there is no \moire-of-\moire pattern to speak of.}
	% \label{mom_1}
	%\end{figure}
	The \moire-of-\moire length scale emerges due to the misalignment of the \moire patterns between layers 1 and 2, and layers 2 and 3, $(\theta_{12}+\theta_{23})/2$, as shown in the Fig. 1(c) of the main text. Starting from a particular momentum point in the second layer, we may reach other points in the second layer by hopping through either the first or third layer. Hoppings between only 2 of the layers allow us to define a momentum space \moire lattice, a subset of which are shown as red points along the black lines. The dotted line marks the closest approach between middle layer states connected by hoppings that only pass through either the first or third layer, which defines the \moire-of-\moire modulation lengthscale. The sketch specifically shows the situation for ${m,n}={3,4}$, $\theta_0=1^\circ$.
	
	The distance between a momentum state reached by $a$ hops through the first layer and another state reached by $b$ hops through the third layer is $\sqrt{3}k_\mathrm{D}P_{ab}$, where
	\begin{align}
		P_{ab}^2 &= (a\theta_{12})^2+(b\theta_{23})^2-2ab\abs{\theta_{12}\theta_{23}}\cos\left(\frac{\theta_{12}+\theta_{23}}{2}\right)\nonumber\\
		&\approx(a\abs{\theta_{12}}-b\abs{\theta_{23}})^2+ab\abs{\theta_{12}\theta_{23}}\left(\frac{\theta_{12}+\theta_{23}}{2}\right)^2\nonumber\\
		&\rightarrow(am-b\abs{n})^2\theta_0^2+abm\abs{n}\left(\frac{m+n}{2}\right)^2\theta_0^4,
		\label{moml}
	\end{align}
	and we take the small angle approximation in the second line and commensurate angles ($m\geq0$ WLOG) in the third line. The \moire-of-\moire lengthscale is then $\l_{m2}=2\pi/\sqrt{3}k_\mathrm{D}\mathrm{min}(P_{ab})$, 
	where we minimise $P_{ab}$ over all integers $a\geq0$ and $b\geq0$, except for the trivial solution $P_{00}=0$. For small commensurate angles with $m$ and $n$ small, this generally occurs at $P_{\abs{n}m}=\frac{\abs{mn(m+n)}}{2}\theta_0^2$, however, if $m$ and $n$ are large and of the same sign, $P_{ab}$ may be minimised at some other $a\neq\abs{n}$, $b\neq m$. For example, at $\theta_{12}=1.0^\circ$, $\theta_{23}=0.9^\circ$, $P_{1,1}<P_{9,10}$. For completeness, we note that this description only considers lattice points lying on the two lines as shown in the sketch, when in fact there are others in the 2D plane. Provided the angles and thus the misalignment is small, the points on these lines remain the closest to each other. 
	
	Numerically calculated results for $l_{m2}$ are shown in the Fig. 1(d) of the main text. As expected, $l_{m2}$ diverges along the line $\theta_{12}=-\theta_{23}$, and is generally larger for $\theta_{23}/\theta_{12}$ negative than for $\theta_{23}/\theta_{12}$ positive due to the reduced misalignment of the \moire patterns. $l_{m2}$ is also particularly pronounced at commensurate angles with simple ratios, where it is easy to minimise the first term of \eqnref{moml} without the second term getting too large. For experimental samples smaller than $l_{m2}$, a particular local Hamiltonian expansion is expected to be a good description of the bands while for samples much larger than $l_{m2}$, it is necessary to average over expansion centers, or equivalently, over $\vect{D}_1-\vect{D}_3$. We note that the form for $l_{m2}$ derived and calculated here agrees with that previously derived~\cite{MGM23} for the special case of $\theta_{23}/\theta_{12}\approx1$.
	
	Neglecting the slow periodicity in the local expansion, as is done in the numerical calculations of us and \cite{MGM23}, neglects the misalignment and so forces the black lines of the upper panel sketch to coincide. In the case of commensurate angles, this leads to $\mathrm{min}(P_{ab})=0$ and thus $l_{m2}$ diverging, leading us to ignore the modulation of $\vect{D}_1-\vect{D}_3$.

	%%%%%%%%%%%%%%%%%%%%%%%%%%%%%%%%%%%%%%%%%%%%%%%%%%%%%
	\section{Finding the magic surface in trilayer configuration space} 
	Our numerical calculations have shown that a magic angle pair may be found seemingly for any numerically accessible (commensurate) twist angle ratio, and so the natural question would be whether this may be extended to the continuum in angle-pair-space. We start by writing the trilayer Hamiltonian in layer space, 
	
	\begin{align}\label{trilayerH}
		H=
		\begin{pmatrix}
			D_1 & {\cal H}_T^{12} & 0 \\
			{\cal H}_T^{12\dagger} & D_2 & {\cal H}_T^{23} \\
			0 & {\cal H}_T^{23\dagger} & D_3
		\end{pmatrix}_{L},
	\end{align}
	with $D_l$ the Dirac cones of layer $l$ and the ${\cal H}_T^{ij}$ as in \eqnref{hamiltd}. The subscript $L$ denotes layer space, and each ``matrix element'' here is itself a matrix in sublattice space. In the chirally symmetric limit, $t_{AA}=0$, a simple permutation allows us to rewrite the Hamiltonian in the dimensionless form 
	
	\begin{align}\label{trilayerHperm}
		H_\mathrm{chiral}/t_{AB}=
		\begin{pmatrix}
			0 & \mathcal{D}^\dagger \\
			\mathcal{D} & 0
		\end{pmatrix}_{\mathrm{AB}},
	\end{align}
	where the subscript AB indicates sublattice space and the zeroes on the diagonals are the natural consequence of using the chirally symmetric model.  The layer space submatrix is given by
	
	\begin{align}\label{chiralsubmat}
		\mathcal{D}=
		\begin{pmatrix}
			-i\frac{\hbar\vF}{t_{AB}}\bar{\partial} & U_{12}(\bfr) & 0 \\
			U_{12}(-\bfr) & -i\frac{\hbar\vF}{t_{AB}}\bar{\partial} & U_{23}(\bfr) \\
			0 & U_{23}(-\bfr) & -i\frac{\hbar\vF}{t_{AB}}\bar{\partial}
		\end{pmatrix}_{L},
	\end{align}
	where $\bar{\partial}_i=\tfrac{1}{2}(\partial_x+i\partial_y)$ and their relative twists have been compensated for by corresponding countertwists of the spinors. $U_{12}(\bfr)=\sum_{j=1,2,3} e^{i \frac{2 \pi ( j- 1)}{3}} e^{i m \vec{g}_j \cdot \left( \vec{r} - \vec{\mathbf{D}}_1 \right)}$ is the \moire potential between layers 1 and 2, and similarly for $U_{23}(\bfr)$ replacing $m$ with $n$ and $\mathbf{D}_1$ with $\mathbf{D}_3$. 
	
	The off-diagonal elements of the layer space submatrix have the form 
	\begin{align}\label{chiralsubmatskel}
		\mathcal{D}_{off}=
		\begin{pmatrix}
			0 & a & 0 \\
			b & 0 & c \\
			0 & d & 0
		\end{pmatrix}_{L},\quad\{a,b,c,d\}\in\mathbb{C},
	\end{align}
	which may be collected into a single off-diagonal through the similarity transformation
	\begin{align}\label{simtrans}
		S^{-1}\mathcal{D}_{off}S=
		\begin{pmatrix}
			0 & 0 & 0 \\
			0 & 0 & a\sqrt{1+\frac{cd}{ab}} \\
			0 & b\sqrt{1+\frac{cd}{ab}} & 0
		\end{pmatrix}_{L^\prime},\quad S=\frac{1}{\sqrt{2}}\begin{pmatrix}
			-\frac{c}{b} & 0 & \frac{a}{d} \\
			0 & \sqrt{2} & 0 \\
			1 & 0 & 1
		\end{pmatrix},
	\end{align}
	where the subscript $L^\prime$ reminds us of the nontrivial mixing of the layers. We are thus able to re-express the interfering double-\moire problem into a single effective \moire potential. We note that the transformation $S$ is not unitary in general and so does not conserve the inner products of spinors, but does conserve the bandstructure. We further note that there is some freedom to choose the form of the single off-diagonal element; our choice most easily connects to previous work~\cite{KKTV19} where in the case of alternating twist angles, $U_{12}(\bfr)=U_{23}(-\bfr)$ and so $S$ is unitary and the form factor $\sqrt{1+\frac{cd}{ab}}=\sqrt{2}$, resulting in a constant scaling of the bilayer \moire potential and thus of the magic angle.
	
	Applying the transformation of \eqnref{simtrans} to \eqnref{chiralsubmat} then allows us to obtain a single effective \moire potential, from which an estimate of the magic angle pair may be made by comparison with the bilayer \moire potential, a nontrivial extension of previous work~\cite{KKTV19} which focused on the special case of alternating twist angle multilayers. 
	
	Our estimate for the magic angle and the error of our estimate is calculated from the average and standard deviation of the form factor $\sqrt{1+\frac{cd}{ab}}$ over space. While the form of the effective \moire potential we have chosen does indeed have singularities e.g. at the origin, these diverge as $\tfrac{1}{x+i y}$ and so the integral over the plane converges, which is unsurprising as it is possible to choose a form of effective \moire potential with no singularities at all, though in that case the analogy to twisted bilayer graphene is weaker.
	
	\begin{figure}[h]
		\centering
		\includegraphics[width=0.7\linewidth]{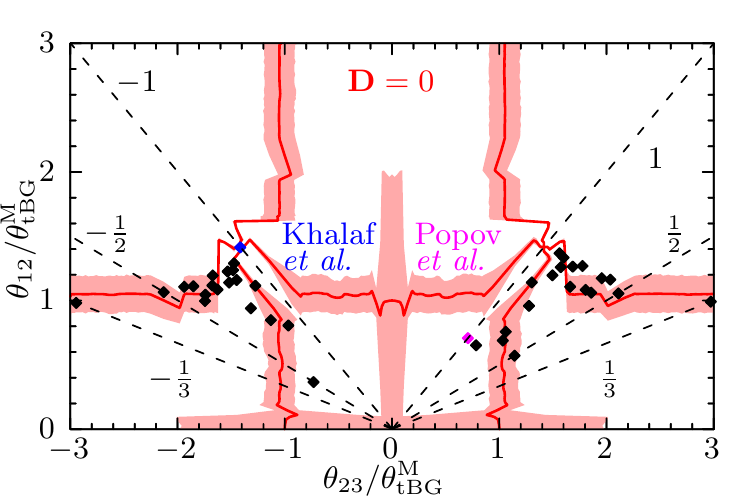}
		\includegraphics[width=0.7\linewidth]{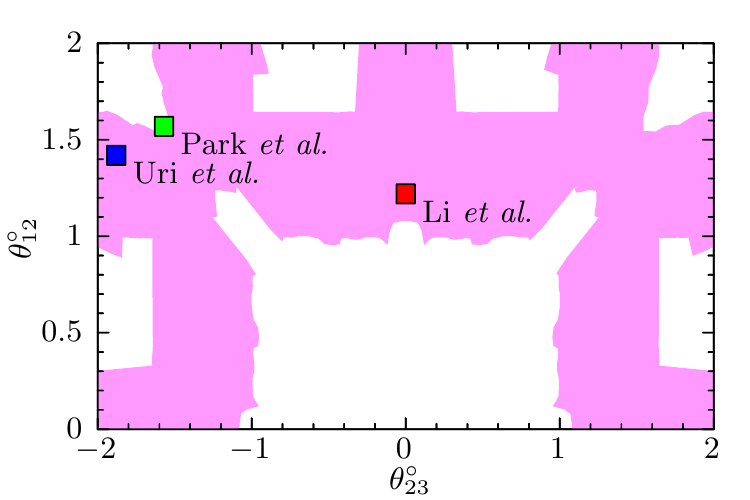}
		\caption{Top: Predicted magic angle pairs $(\theta_{12}^{\mathrm{M}},\theta_{23}^{\mathrm{M}})$ for $\vect{D}=0$. Shaded regions denote the error in the magic angle pair prediction. The configurations studied in~\cite{PT23} and~\cite{KKTV19} are indicated with a black and purple diamond respectively, and those corresponding to our numerical calculations are marked with black diamonds. Dotted lines of constant $\theta_{12}/\theta_{23}$ are marked as guides to the eye. Bottom: Predicted range of magic angles over all $\vect{D}$. The shaded region indicates pairs of angles for which there exists some $\vect{D}$ that would give flat bands. The configurations studied in~\cite{sc_tng,AAtrilayer,mit_exp} are marked in green, red and blue respectively.}
		\label{magang}
	\end{figure}
	
	Results are shown in the upper panel of \figref{magang} for zero shift, $\vect{D}=0$, where we indicate with a red line the predicted magic angle pairs according to the average-value analysis described above, with the uncertainty in the prediction shown by the shaded pink region. The alternating twist case previously studied~\cite{KKTV19} is marked with a blue diamond, and we note that this is the only point where the analysis is exact. We mark also our numerically calculated magic angle pairs with black diamonds, and note the excellent agreement with the more general analysis, giving us confidence that the results are reliable even for the numerically inaccessible case of incommensurate angles. Our analysis predicts two branches of magic angle pairs from which an infinite set descends, as in the bilayer case~\cite{TKV19}. At most two sets of numerically calculated magic angles per angle ratio are shown to connect with the two branches, though other values have been found.
	
	We repeat the calculation over all \moire shifts to obtain a range of magic angle pairs, shown in the lower panel of \figref{magang}.  The magic angles vary continuously with shift and so the manifold of magic configurations is a three-dimensional surface within the four-dimensional trilayer configuration space of two angles and a two-dimensional relative shift. This surface divides the configuration space into unconnected sectors, as seen in the lower panel of~\figref{magang} where it is not possible to move from $\theta_{12}=\theta_{23}=0$ to $\theta_{12}=\theta_{23}=2^\circ$ without passing through the magic surface.  We remind the reader here that the plot is symmetric about the origin. This may have consequences on the topology of the bands as a function of trilayer configuration, as magic configurations may arise as the ``midpoints'' of band crossings, with the flat bands flattened by repulsion of the crossing bands. Previous work on alternating-twist magic angle trilayer graphene~\cite{sc_tng}, with trilayer configuration $(\theta_{12}=1.57^\circ,\theta_{23}=1.57^\circ,\vect{D}=0)$, a twisted monolayer--$AA$-bilayer system~\cite{AAtrilayer}, configuration $(\theta_{12}=1.22^\circ,\theta_{23}=0,\vect{D}=0)$, and a \moire quasiperiodic crystal~\cite{mit_exp}, with configuration $(\theta_{12}=1.41^\circ,\theta_{23}=1.88^\circ,\vect{D}\:\mathrm{unknown})$, may be placed as single points within this extended magic phase.

	%%%%%%%%%%%%%%%%%%%%%%%%%%%%%%%%%%%%%%%%%%%%%%
	\section{Tight-binding results} 
\begin{figure}[t]
		\includegraphics[width=0.6\textwidth]{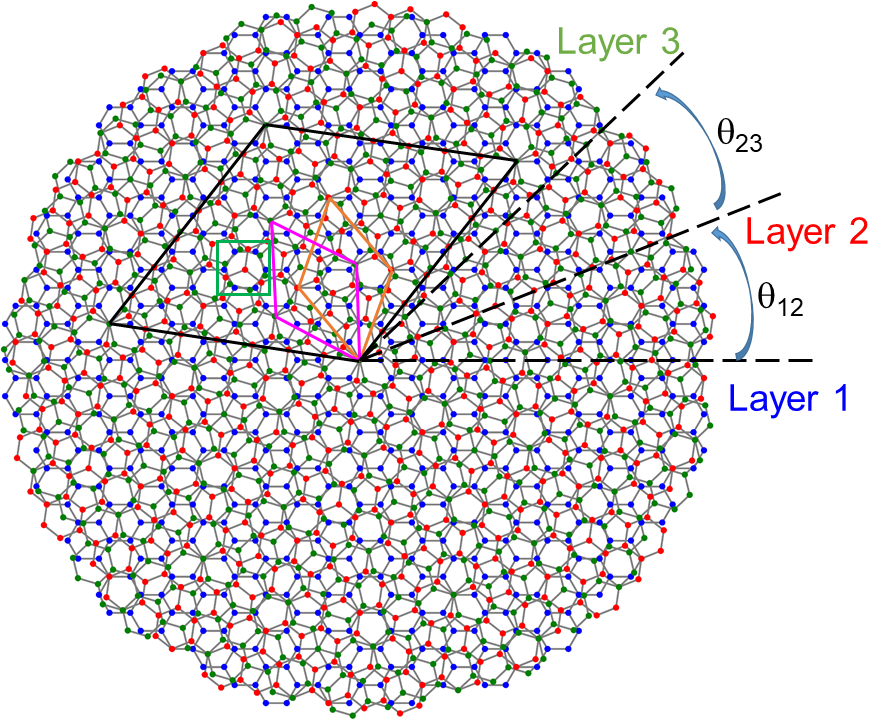}
		\caption{Schematic of the $AAA$ trilayer graphene with $\theta_{12}=\theta_{23}=\theta_0=21.8^\circ$. The $\theta_{12}$ and $\theta_{23}$ are the twist angles between layer 1 and layer 2, layer 2 and layer 3, respectively. The orange and purple rhombi are the moir\'e unit cells with $\theta_{12}$ and $\theta_{23}$, respectively, which has length $\ell_m=0.65$ nm. The black rhombus is the unit cell of the moir\'e-of-moir\'e with $\ell_{m2}=1.7$ nm. The green square shows an $ABA$ stacking region.}
		\label{fig:structure}
	\end{figure}

	We use a round disk method to construct the twisted graphene trilayers with arbitrary twist angles. The two independent twist angles $\theta_{12}$ and $\theta_{23}$ are chosen to be the rotation of the second layer 2 relative to the first layer 1 and the rotation of the third layer 3 relative to the second layer 2, respectively. The rotation origin is chosen at an atom site. We use a twist angel pair ($\theta_{12}$, $\theta_{23}$) as the notation for different twist angle configurations. Positive (negative) values of the twist angle denotes counterclockwise (clockwise) rotations. The sample with (-$\theta$, $\theta$) has a mirror symmetry with the middle layer as the mirror plane. To calculate the property of these large scale systems with arbitrary twist angles, we construct the system in a large round disk. The radius of the disk should be set sufficiently large to rid the effects of edge states~\cite{yu2019dodecagonal}. In the actual calculation, the disk with radius of $172.2$ nm ($700d$ being $d$ the lattice constant of graphene) and contains 10 million carbon atoms are utilized for the twist angles investigated in this work.  Figure \ref{fig:structure} shows the ($21.8^\circ$, $21.8^\circ$) configuration of twisted trilayer graphene with $AAA$ stacking.  
	
	A parameterized full tight-binding (TB) model is used. In the tight-binding model, only $p_z$ orbitals are taken into account, we construct the Hamiltonian of the twisted graphene trilayers as
	\begin{equation}\label{Hamil}
		H = \sum_i\epsilon_i |i \rangle\langle i|+\sum_{\langle i,j\rangle}t_{ij} |i \rangle\langle j|,
	\end{equation} 
	where $|i\rangle$ is the $p_z$ orbital located at $\mathbf{r}_{i}$, and $\langle i,j\rangle$ is the sum over index $i$ and $j$ with $i \neq j$.  
	According to the Slater-Koster (SK) formalism, the hopping integral $t_{ij}$, interaction between two $p_z$ orbitals located at $\mathbf{r}_{i}$ and $\mathbf{r}_{j}$ has the form~\cite{SK1954simplified}
	\begin{equation}
		t_{ij}=n^2V_{pp\sigma}(r_{ij})+(1-n^2)V_{pp\pi}(r_{ij}),
	\end{equation}
	where $r_{ij}=|\mathbf{r}_{j}-\mathbf{r}_{i}|$ is the distance between $i$ and $j$ sites, with $n$ as the direction cosine along
	the direction $\bm{e_z}$ perpendicular to the graphene layer . The Slater and Koster parameters $V_{pp\pi}$ and $V_{pp\sigma}$ follow
	\begin{equation}
		\begin{aligned}
			V_{pp\pi}(r_{ij})=-t_0e^{q_\pi(1-r_{ij}/a)}F_c(r_{ij}),\\
			V_{pp\sigma}(r_{ij})=t_1e^{q_\sigma(1-r_{ij}/h)}F_c(r_{ij}),
		\end{aligned}
	\end{equation}
	where $a=1.42$ \AA \; and $h=3.349$ \AA \; are the nearest in-plane and out-of-plane carbon-carbon distance, respectively, $t_0=3.1$ eV and $t_1=0.43$ eV are the TB hopping parameters taken from the moir\'e quasiperiodic crystal paper \cite{mit_exp}. The parameters $q_\sigma$ and $q_\pi$ satisfy $\frac{q_\sigma}{h}=\frac{q_\pi}{a}=2.218$ \AA$^{-1}$, and the smooth function is $F_c(r)=(1+e^{(r-r_c)/l_c})^{-1}$, in which $l_c$ and $r_c$ are chosen as $0.265$ and 5.0 \AA, respectively. In the twisted graphene trilayer calculations, we only consider the interlayer hoppings between adjacent layers. An open boundary condition is used in the disk model. The electronic properties of the twisted graphene trilayer with arbitry angles ($\theta_{12}$,$\theta_{23}$) are calculated by the tight-binding propagation method implemented in the TBPLaS simulator \cite{li2023tbplas}. For example, the detailed formula of the density of states is
	\begin{equation}
		\label{dos}
		D(\varepsilon)=\frac{1}{2\pi S}\displaystyle\sum_{p=1}^{S}\int_{-\infty}^{\infty}e^{i\varepsilon t}\langle\varphi_p(0)|e^{-iHt}|\varphi_p(0)\rangle dt,
	\end{equation}
	where $|\varphi_p(0)\rangle$ is one initial state which is the random superposition of all basis states, %$p$ stands for different random initial states. 
	$S$ is the number of random initial states. The distribution of states in real space can be obtained by calculating the quasieigenstates \cite{li2023tbplas} (a superposition of degenerate eigenstates with certain energy). The quasieigenstates has the expression:
	\begin{equation}\label{quasi}
		|\Psi(\varepsilon)\rangle=\frac{1}{\sqrt{\sum_n|A_n|^2\delta(\varepsilon-E_n)}}\sum_nA_n\delta(\varepsilon-E_n)|n\rangle,
	\end{equation} 
	where $A_n$ are random complex numbers with $\sum_{n}|A_n|^2=1$, $E_n$ is the eigenvalue and $|n\rangle$ is the corresponding eigenstate. The local density of states (LDOS) mapping calculated from the quasieigenstates is highly consistent with the experimentally scanning tunneling microscopy $\mathrm dI/\mathrm dV$ mapping. We also consider the lattice relaxation effect on the electronic properties of twisted trilayer graphene. In the round disk sample, the edge carbon atoms possessing dangling bond are passivated by placing in-plane hydrogen atoms to saturate the dangling $\sigma$ edge bonds. The carbon-hydrogen bond length is assumed to be 0.1 nm. We employ the classical molecular dynamics simulation package LAMMPS to do the full (both in-plane and out-of-plane) lattice relaxation \cite{thompson2022lammps}. The intralayer C-C and C-H interactions and interlayer C-C interactions are simulated with REBO \cite{brenner2002second} and kolmogorov/crespi/z version of Kolmogorov-Crespi (KC) \cite{kolmogorov2005registry} potentials, respectively. 
	
	\begin{figure}[t!]
		\includegraphics[width=0.5\textwidth]{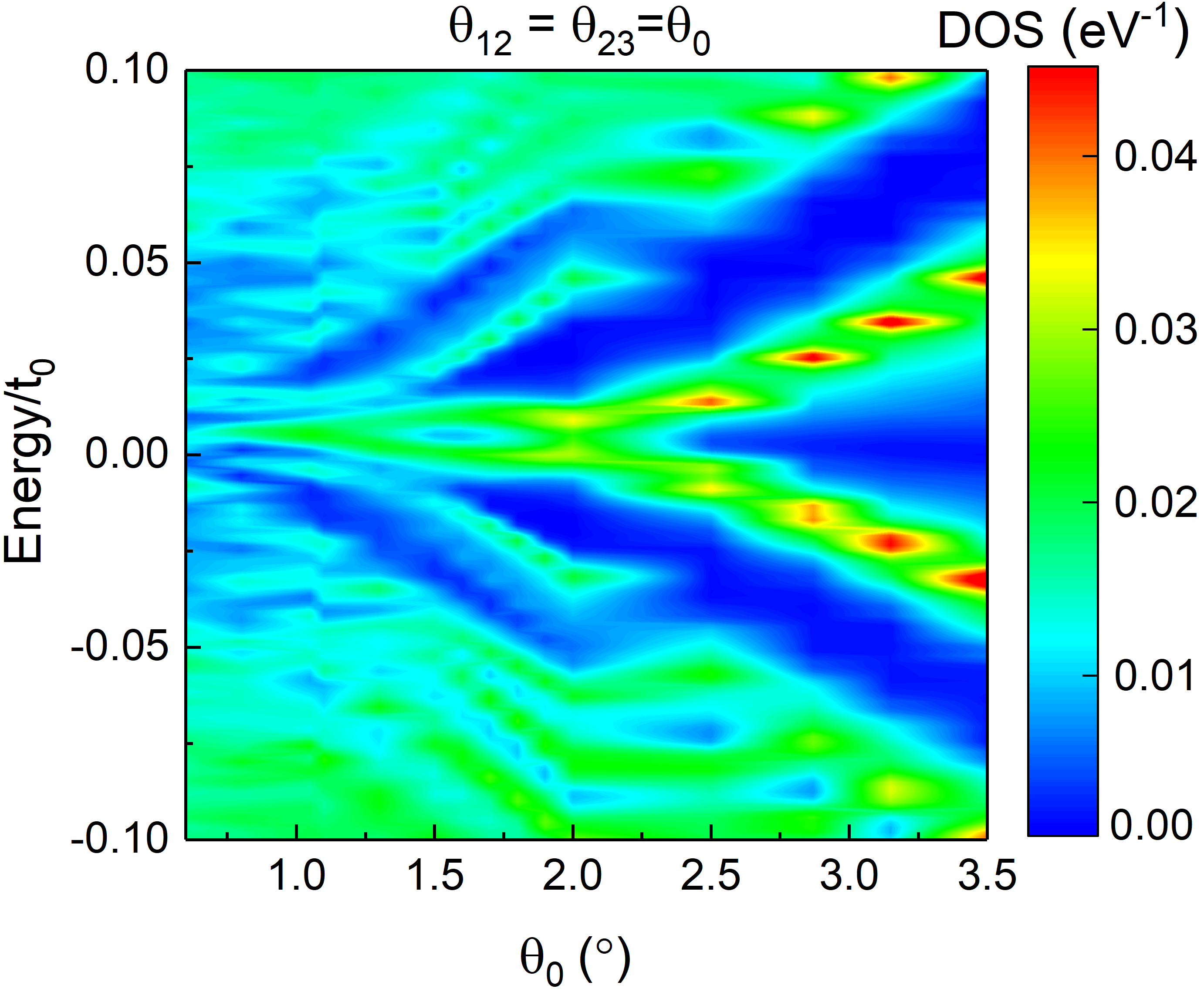}
		\caption{The TB DOS of $AAB$ trilayer graphene with twist angles $\theta_{12}=\theta_{23}=\theta_0$. The hopping parameters are $t_0=2.8$ eV and $t_1=0.44$ eV. The displacement $\mathbf{D}$ is zero. }
		\label{fig8:dos}
	\end{figure} 
	
\begin{figure}[t!]
		\includegraphics[width=0.6\textwidth]{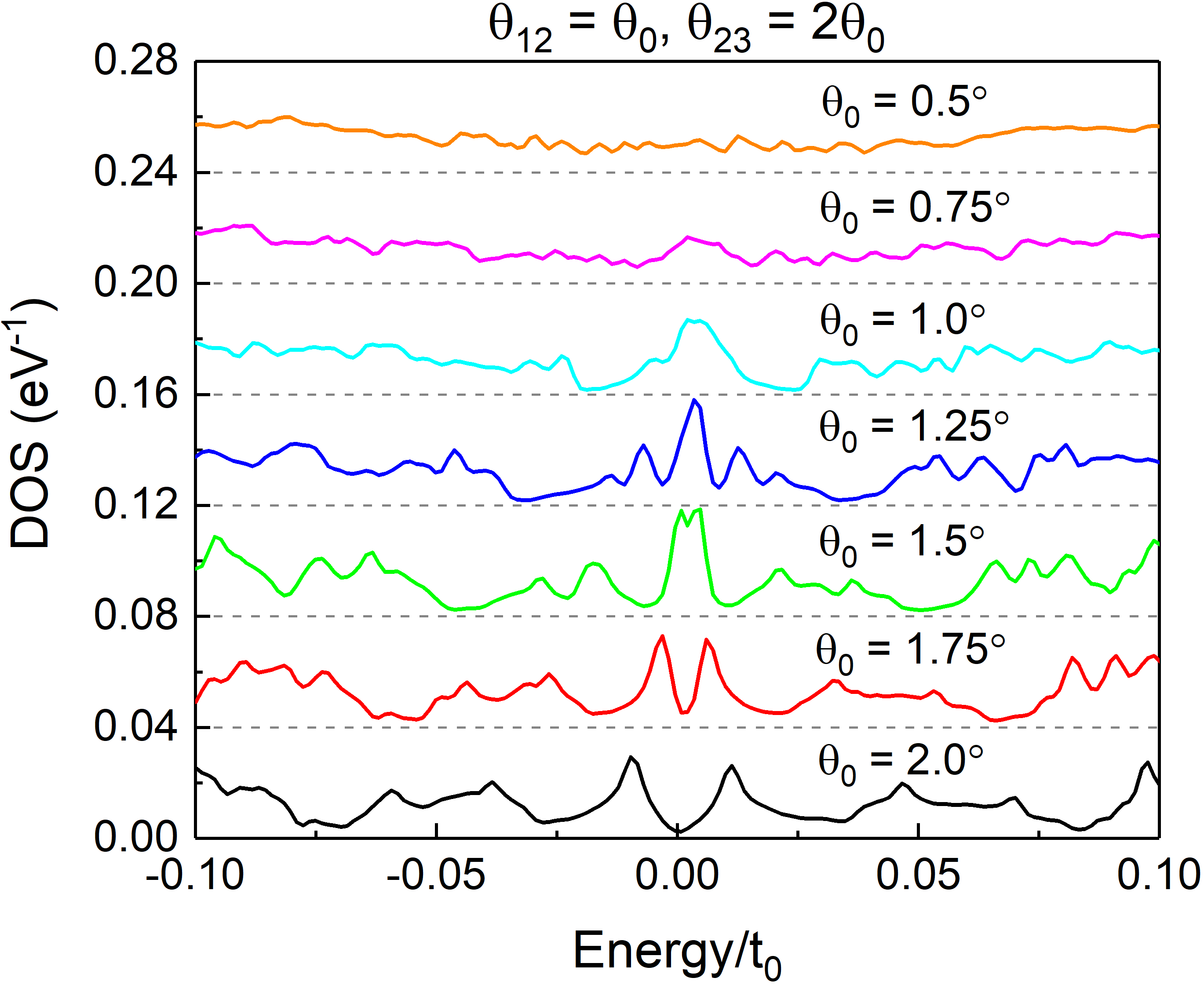}
		\caption{The TB DOS of $AAB$ trilayer graphene with $\theta_0$ where twist angles $\theta_{12}=\theta_0$, $\theta_{23}=2\theta_0$. The hopping parameters are $t_0=2.8$ eV and $t_1=0.44$ eV. The displacement $\mathbf{D}$ is zero. Curves are vertically shifted for clarity.}
		\label{fig9:dos}
	\end{figure}

	We focus on the helical twisted trilayer graphene configuration. Firstly, we consider the particular case that the two twist angles are equal. In the TB calculation, we set $t_0=2.8$ eV and $t_1=0.44$ eV, which give a magic angle of $1.05^\circ$ in the twisted bilayer graphene case. The evolution of the density of states with twist angles of $AAB$ twisted trilayer graphene is shown in Fig. \ref{fig8:dos}. Note the discontinuity of the van Hove singularity (VHS) evolves with the twist angle is due to the low resolution of the twist angle, which will not change our conclusion. The system can be considered as a $AAA$ trilayer with a displacement of the carbon-carbon distance $a$ between the top and bottom layers. The tendency is similar to the $AAA$ trilayer graphene case \cite{MZY23,ZCMLK20}. That is, with $\theta_0$ decreases, the van Hove singularity gap (the energy difference between the first van Hove singularity on the valence and conduction bands) narrows and reaches minimum value at around $2^\circ$. This result is reasonable since the moir\'e unit cell with periodicity $\ell_{m2}$ contains all the configurations, for instance, the $AAA$ and $ABA$ stackings. In this case, the first "magic angle" is $2^\circ$. The VHSs merge again at the angle $\theta_0=1^\circ$. In the magic angle of trilayer in chiral limit in Fig. (3) of the main text, there is a magic angle at $\theta_0=0.72^\circ$. In the case of $\theta_{12}=\theta_0$, $\theta_{23}=2\theta_0$, shown in Fig. \ref{fig9:dos}, the first magic angle is $\theta_0=1.25^\circ$, which is consistent with the result in the continuum limit in Ref. \cite{zhu2020twisted}.  
	
	\begin{figure}[h!]
		\includegraphics[width=0.9\textwidth]{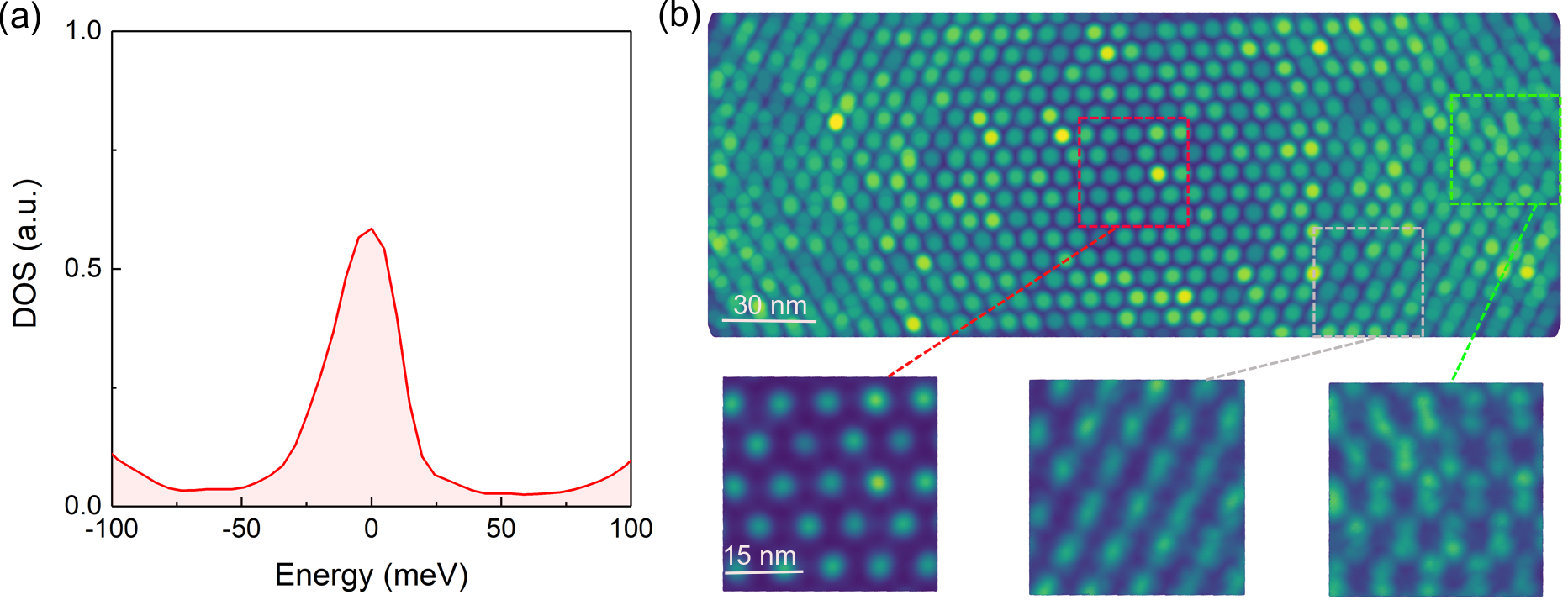}
		\caption{The lattice relaxation effect. (a) The TB DOS of $AAA$ trilayer graphene with magic twist pair $\theta_{12}=\theta_{23}=\theta_0=1.57^\circ$. (b) The local DOS at zero energy in real space with a zoom-in of the states at different displacement $\mathbf{D}$. The colors of the squares correspond to the ones in Fig. 2(a) of the main text.  The hopping parameters are $t_0=2.8$ eV and $t_1=0.44$ eV.}
		\label{fig10:relax}
	\end{figure} 
	
	As discussed in the Ref. \cite{MZY23}, in the case of $AAA$ twisted trilayer graphene with $\theta_{12}=\theta_{23}$, the first magic angle decreases from $2^\circ$ to $1.57^\circ$ when the lattice relaxation is considered. A sharp peak is located at the charge neutrility point, shown in Fig. \ref{fig10:relax}(a). The states of the peak in the real space are different for different displacement $\mathbf{D}$ are different, which is in agreement with the result in Fig. 2(a) of the main text. Such state feature can be detected by the scanning tunneling microscopy in experiment. However, For the relaxation of the system with the REBO and KC potentials, we could not find a strong reconstruction of the structure with the $ABA$ region relax to large area and clear domain wall region, which may achieve via a stronger intralyer and interlayer potential \cite{Detal23}.
	
	\section{Dielectric screening in trilayers}
	
	It has been widely shown that the dielectric response for multi-twisted materials increases drastically as the bands become flat \cite{Goodwin2019Attractive}. The plasmon screening on electron-electron interaction is believed to have a significant effect on the strong correlation phenomena. One potential advantage of magic multilayers over tBG is that the calculated noninteracting bandstructures are expected to be more resilient against Hartree effects due to the higher dielectric function from plasmon screening.
	
	We may verify this by calculating the static component of the dielectric function. For simplicity, we ignore Umklapp processes, which would give a matrix structure to the polarization, with entries labelled by moiré reciprocal lattice vectors, $\vec{G}, \vec{G}'$. Within the framework of random phase approximation (RPA), the dynamic dielectric function can be calculated via
	\begin{equation}
		\epsilon(\mathbf{k}, \omega)=1-V(\mathbf{k})\Pi(\mathbf{k}, \omega)
	\end{equation}
	where $V(\mathbf{k})=2\pi e ^2/\kappa k $ is the Fourier component of the two-dimensional Coulomb interaction, with $\kappa$ being the background dielectric constant. We choose $\kappa=4$ in our calculation, in effect assuming the substrate is hexagonal boron nitride (hBN). 
	$\Pi(\mathbf{k}, \omega)$ is the dynamic polarization function given by
	\begin{equation}
		\Pi(\mathbf{k}, \omega)=2 \sum_{\mathbf{q},\xi} \sum_{m, n} \frac{\left(f_{\mathbf{q}+\mathbf{k},\xi}^n-f_{\mathbf{q},\xi}^m\right) F_{\mathbf{q}, \mathbf{q}+\mathbf{k},\xi}^{n m}}{E_{\mathbf{q}+\mathbf{k},\xi}^n-E_{\mathbf{q},\xi}^m-\omega-i0}
		\label{polarization}
	\end{equation}
	here $m,n$ are band index, $\xi$ is valley index, $f_{\mathbf{q}}^m$ is the Fermi-Dirac distribution, and $F_{\mathbf{q}, \mathbf{q}+\mathbf{k},\xi}^{n m}=\left|\psi_{n, \mathbf{q}+\mathbf{k},\xi}^{\dagger} \psi_{m, \mathbf{q},\xi}\right|^2$ is the form factor of two different Bloch states. 
	
	\begin{figure}[h!]
		\includegraphics[width=0.4\textwidth]{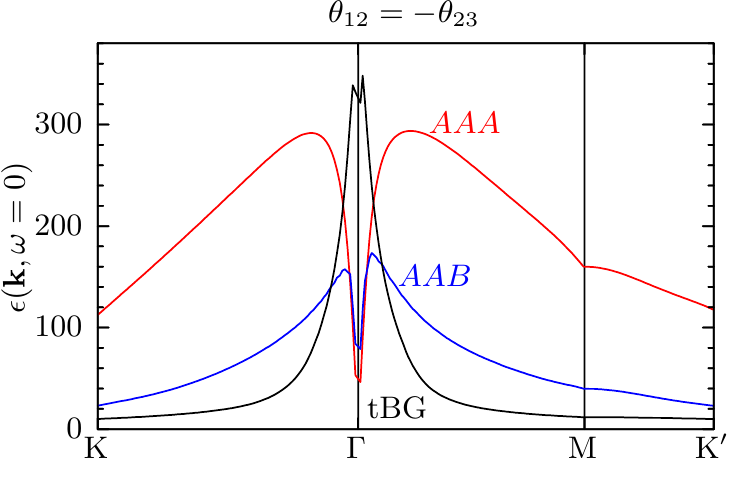}
		\caption{Static component of the dielectric function, $\epsilon(\mathbf{k},\omega=0)$ evaluated over the Brillouin zone for alternating-twist magic angle trilayer graphene at $AAA$ (\textcolor{red}{red}) and $AAB$ (\textcolor{blue}{blue}) stacking, compared to that of magic angle tBG (black). The value of $\epsilon$ for trilayers is greatly increased compared to that of tBG everywhere except the $\Gamma$ point, indicating an overall greater degree of plasmon screening in trilayers and thus greater robustness of the calculated noninteracting bandstructures to Hartree effects.}
		\label{fig11:dielectric}
	\end{figure} 
	
	Results are shown in Fig. \ref{fig11:dielectric}, which shows $\epsilon(\bfk,\omega=0)$ over the \moire Brillouin zone for alternating-twist magic angle trilayer graphene in the $AAA$ and $AAB$ configurations, and for tBG. We choose to compare with alternating twist trilayers to remove \moire-of-\moire effects, and to be able to make direct comparisons between the Brillouin zones of the two materials. The value of $\epsilon$ is seen to be $\mathbf{D}$ dependent, as expected from the $\mathbf{D}$ dependance of the bandstructure, and is noticeably higher than that of tBG everywhere except at $\Gamma$, indicating that there is a greater degree of plasmon screening in the trilayer, as expected, and that the calculated noninteracting bands are therefore expected to be more robust against Coulomb interaction effects than those of tBG.
	
	\section{Electronic bands and Berry curvature.}
\begin{figure}[h!]
		\includegraphics[width=0.3\textwidth]{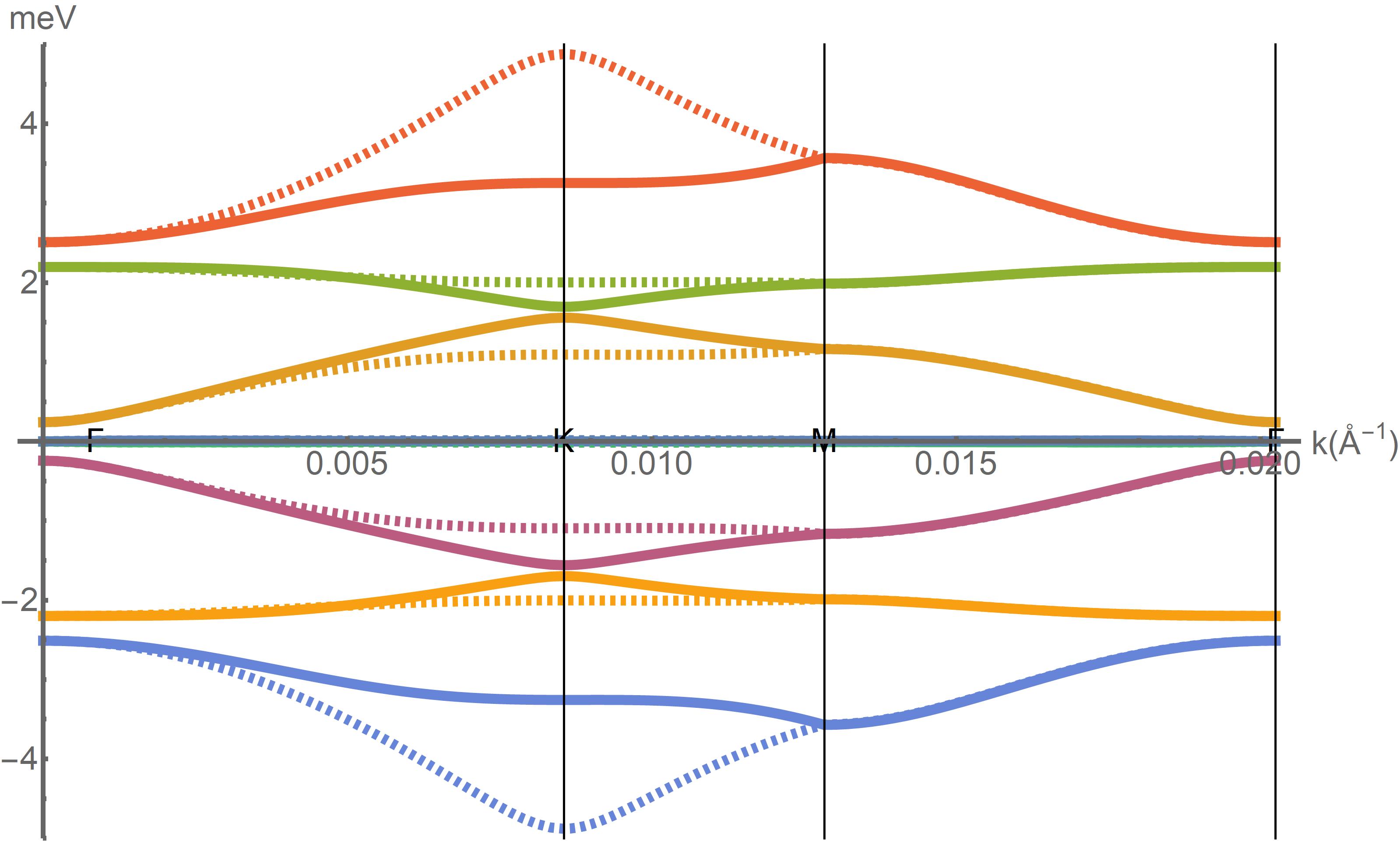}
		\\
		\includegraphics[width=0.8\textwidth]{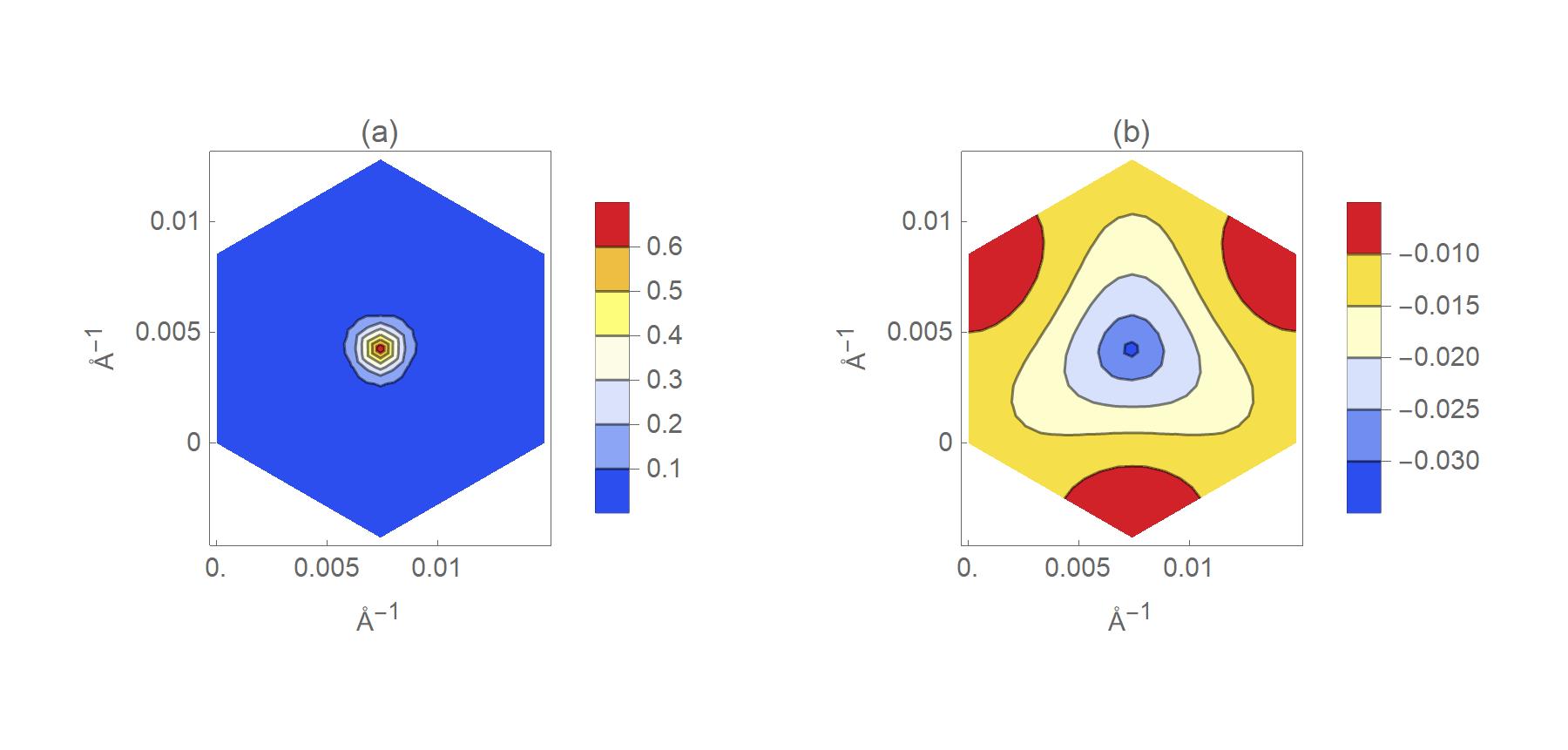}
		\caption{Top: Low energy electronic bands for an $ABA$ trilayer with $\{ m , n \} = \{ 3 , -4 \}$ and $\{ \theta_{12} , \theta_{23} \} = \{ 0.85^\circ , -1.14^\circ \}$. Using the parameters described in the main text, two flat bands are obtained at $\epsilon = 0$. Top: Low energy electronic bands. Bottom: Berry curvature normalized to $2 \pi {\cal C}$ when integrated over the Brillouin Zone, where ${\cal C} = \{ 2 , -1 \}$ is the Chern number of each band.}
		\label{fig:berry}
	\end{figure}

\begin{figure}[h!]
		\includegraphics[width=0.3\textwidth]{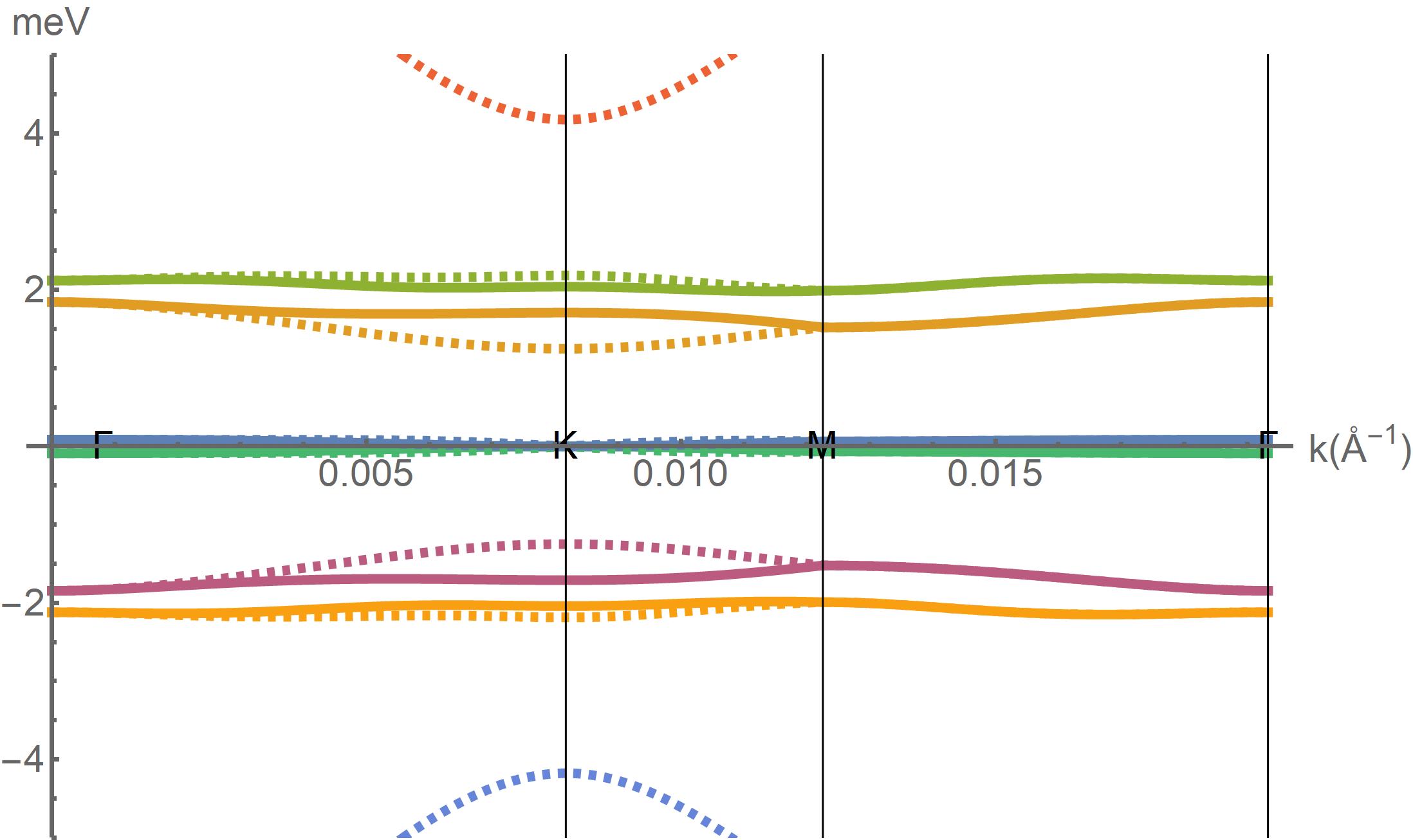}
		\\
		\includegraphics[width=0.8\textwidth]{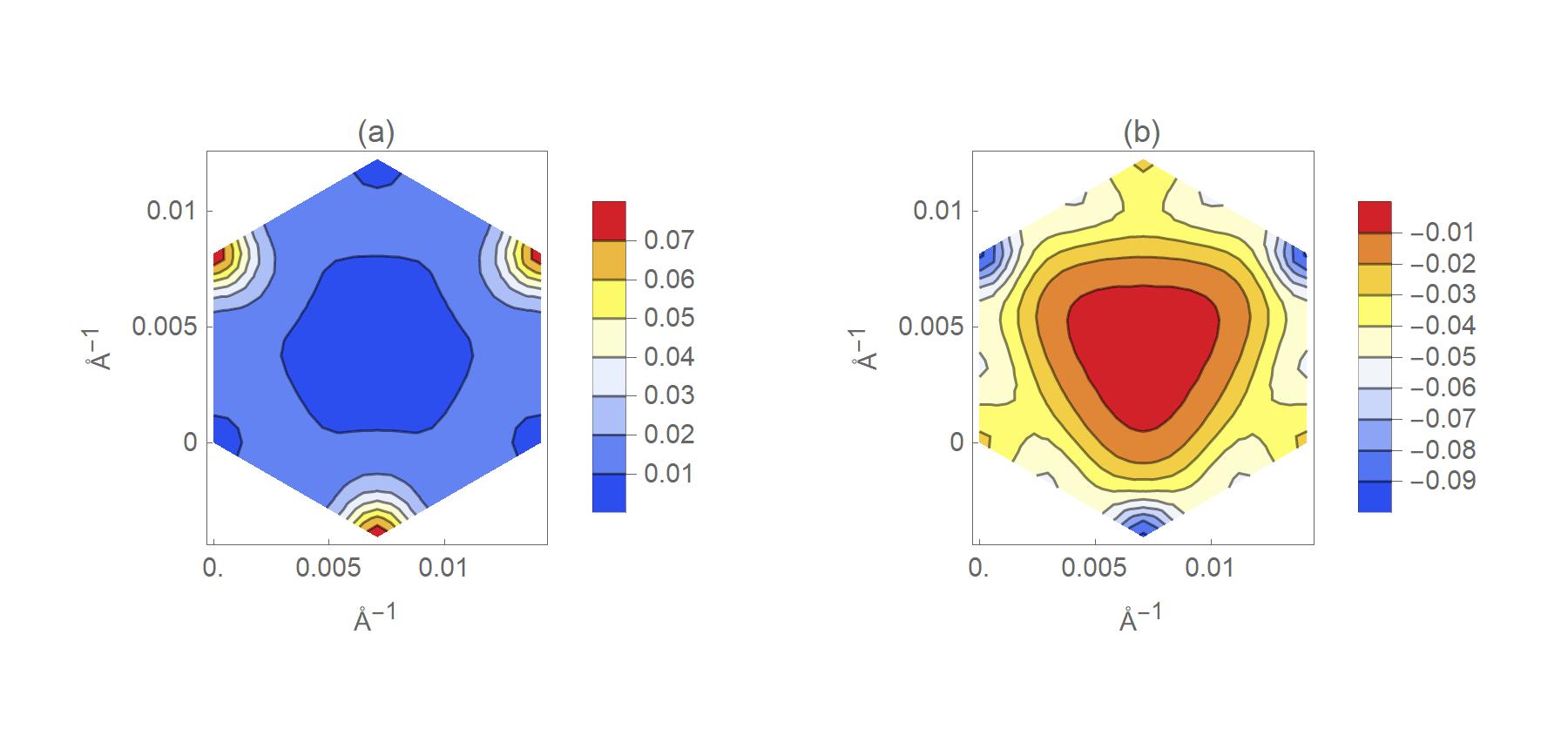}
		\caption{As in Fig.[\ref{fig:berry}] but for $\{ m , n \} = \{ 2 , -3 \}$ and $\{ \theta_{12} , \theta_{23} \} = \{ 0.54^\circ , -0.81^\circ \}$. The Chern numbers are ${\cal C} = \{ 1 , -2 \}$.}
		\label{fig:berry1}
	\end{figure}

	The flat electronic bands at magic angles studied in the main text show finite Berry curvatures, and, generally, non zero Chern numbers, see also~\cite{GMM23,Detal23}. Examples are given in given in Fig.[\ref{fig:berry}] and Fig.[\ref{fig:berry1}].

	\bibliography{biblio2}
\end{document}